\documentclass[onecolumn]{aa} % for a paper on 1 column
\usepackage{graphicx}
\usepackage{multirow}
\usepackage{longtable}
\usepackage{epstopdf}
\usepackage{ulem}
\usepackage{bm}
\usepackage{subfigure}
\usepackage{array}
\usepackage{color}
\usepackage{graphicx}
\usepackage{txfonts}
\begin{document}
\def\apj{{ApJ}}
\def\mnras{{MNRAS}}
\def\aa{{A\&A}}
\def\Nature{{Nature}}
\def\GCN{{GCN Circ}}
\def\PRD{{Phys. Rev. D}}
\def\PRL{{Phys. Rev. Lett}}
\def\etal{{\it et al.}}
\def\gr{{$\gamma$-ray}}

   \title{Searching for an additional high-energy component in Fermi-LAT GRB afterglows}

   \author{Xin-Bo He \and Pak-Hin Thomas Tam \and Guang-Bo Long \and Partha Sarathi Pal \and Yong Zhang \and Li-Jun Zhang}

%   \author{He Xin-Bo\inst{1} , Tam Pak-Hin Thomas\inst{1}, Long Guang-Bo\inst{1} , %Partha Sarathi Pal\inst{1} \and   Zhang Yong\inst{1},}

   \institute{School of Physics and Astronomy, Sun Yat-sen University, Zhuhai 519082, P. R. China,
            \\ \email{tanbxuan@mail.sysu.edu.cn,zhangyong5@mail.sysu.edu.cn}}

   \date{Received XXX; accepted XXX}

% \abstract{}{}{}{}{}
% 5 {} token are mandatory

  \abstract
{The very high-energy (VHE;  $\ge$ 100 GeV) component from at least two gamma-ray bursts (GRBs), i.e., GRB 180720B and GRB 190114C, has been detected in the afterglow phase. It is widely discussed that the GeV to TeV emission is originated from synchrotron self-Compton (SSC) process. The VHE component may cause an upturn at the high-energy spectral ends in the Fermi-Large Area Telescope (Fermi-LAT) observing band.
}
   { We aim to find out whether an additional high-energy component commonly exists in the afterglows of Fermi-LAT GRBs. This study will help us to better understand how common it is for a GRB afterglow detected by Fermi-LAT to involve a VHE component.
    }
   {First, we selected those GRBs that emit $\geq$10 GeV photons. The $\geq$10~GeV photons can be considered as a plausible proxy for a VHE component. We systematically analyzed 199 GRBs detected by Fermi-LAT during 2008--2019. If an additional high-energy component exists in the afterglows of Fermi-LAT GRBs, the best-fit spectral model could be a broken power-law (BPL) model with an upturn above a break energy. We compare the afterglow spectra using power-law (PL) and BPL representations.}
   {Out of the 30 GRBs with $\geq$10~GeV photons that arrived after $T_\mathrm{90}$ (the time duration when 90\% of the prompt emission was detected), 25 GRBs are tentatively or significantly detected at 0.1--200 GeV after 2$\times\,T_\mathrm{90}$. The spectrum of GRB 131231A shows an upturn above an energy break of 1.6$\pm$0.8~GeV, supporting the BPL model. For GRB~131231A, we performed a modeling of its X-ray and $\gamma$-ray spectra, and found that the SSC model can explain the upturn with acceptable parameter values. In the cases of GRB 190114C, GRB 171210A, GRB 150902A, GRB 130907A, GRB 130427A, and GRB 090902B, the improvement of the BPL fit compared to the PL fit is tentative or marginal.
   }
  % Conclusions heading (mandatory)
  {There is no conclusive evidence that an additional higher energy component commonly exists in Fermi-LAT GRB afterglows, except for a group of Fermi-LAT GRBs mentioned above. Such an additional high-energy component may be explained by the synchrotron self-Compton mechanism. Current and future VHE observations will provide important constraints on the issue.}
   \keywords{gamma-ray bursts; radiative mechanisms}
 \authorrunning{He et al.} \titlerunning{On LAT GRB afterglow spectra}

    \maketitle
%
%-------------------------------------------------------------------

\section{Introduction}
\label{sect1}
Gamma-ray bursts (GRBs) consist of short (tens of milliseconds to thousands of seconds) and bright prompt emission at $\sim$keV to MeV energies~\citep[e.g.,][]{Ree92,Meszaros97}, followed by longer (days to months)  afterglow phase at the frequency range from radio to GeV $\gamma$-rays~\citep[e.g.,][]{Costa97}. Generally, the prompt emission is expected to be produced by some internal dissipation mechanisms such as collisionless shock in jet~\citep{Ree94}, and the afterglow emission, from radio up to X-rays, can be described by synchrotron emission of external shock with surrounding medium~\citep{Meszaros97}.

 {\it Fermi Gamma-ray space Telescope\footnote{\url{https://fermi.gsfc.nasa.gov}}}~\citep{Fermi}, launched in 2008, plays an important role in the detection, as well as the spectral and timing characterisation of GRBs, with its two onboard instruments: the Large Area Telescope (LAT, 20 MeV to $>$300 GeV)~\citep{Fermi-LAT} and Gamma-ray  Burst  Monitor (GBM, 8 keV to 40 MeV)~\citep{Fermi-GBM}. At $>$30~MeV, the Fermi-LAT can monitor GRBs in both the prompt and afterglow phases. It has detected 186 GRBs in the first decade at 100 MeV energies~\citep[see, e.g.][]{lat_grb_cat2}; this corresponds to 8$\%$ of all GRBs that are detected with GBM. Photons of $\ga$10 GeV have been detected from GRBs 080916C~\citep{Abdo2009}, 090902B~\citep{Abdo2009A,Panaitescu2017},  090328~\citep{Panaitescu2017}, 090926A~\citep{Swenson,Ackermann2011,Yassine2017}, 100414~\citep{Panaitescu2017}, 110721~\citep{Axelsson,Panaitescu2017}, 110731~\citep{Ackermann,Panaitescu2017}, 130427A~\citep{Tam2013, 130427A}, 130907A~\citep{Tang2014}, 131231A~\citep{Liu2014}, 140619B~\citep{Ruffini,Panaitescu2017}, and 160509A~\citep{Laskar,Tam2017}.

Furthermore, Cherenkov telescope arrays MAGIC and H.E.S.S. have recently detected very high-energy (VHE; $\ga$100~GeV) emission from several GRBs, largely at the afterglow phase: GRB 180720B~\citep{HESS}, GRB 190114C~\citep{MagicA}, and GRB 190829A~\citep{GCN25566}). In the case of GRB~190114C, the highest observed photon energy reach $\sim$1~TeV~\citep{MagicA}.

Concerning the origin of the afterglow GeV emission, synchrotron emission from external shock electrons becomes the ‘standard’ radiation mechanism in the Fermi era but there exists a maximum synchrotron energy, typically $\la$10 GeV~\citep[e.g.,][]{KBD10, PiranNakar10}. It is a great challenge for the traditional synchrotron mechanism to explain the $\ga$10 GeV emission, since the flux emitted by this mechanism above several GeV (the maximal cutoff energy) should be very small. The synchrotron self-Compton (SSC) component is expected to dominate in the GeV-TeV energy range~\citep{Meszaros2000,Sari2001,Zhang2001}, implying that the spectra of such GRBs should have two peaks in the broad-band spectral energy distribution (SED): one at the X-ray band, and the other at the TeV \gr~band. Prompted by the recent VHE detections of GRBs, attempts to explain the VHE emission by the Compton processes have been performed~\citep[e.g.,][]{Wang2019,Fraija2019,Zhang2020}.

Generally, the best-fit spectral model of LAT afterglow spectra is usually a power-law model~\citep[e.g.,][]{lat_grb_cat2}. \citet{Tam2013} was able to identify an additional hard component in the LAT afterglow in GRB~130427A. Following this, \citet{Panaitescu2017} found a sample of LAT GRBs whose afterglow spectrum is better described by a broken power-law (BPL) with a smaller high-energy photon index ($\beta$) than low-energy photon index ($\alpha$), providing a strong evidence of hardening spectrum above a few GeV. This indicates that an inverse Compton spectral component may exist in the GRB afterglow phase, and in particular in the Fermi-LAT energy band. It should be noted that the analysis by \citet{Panaitescu2017} is based on aperture photometry, and is focused on the first 1000s after the burst onsets.

Given the detection of VHE photons by Cherenkov telescope arrays from GRB afterglows up to a day, it is desirable to look for photons with the highest energies accessible to Fermi-LAT, e.g., 10--200~GeV, up to one day post-burst. Indeed, Fermi-LAT covers the spectral range between the two peaks (synchrotron and inverse-Compton) expected in the afterglow spectra, and has likely been seen in GRB 190114C~\citep{MagicB}. In this work, we expand the above works to all LAT-detected GRBs, and up to one day after the burst, using maximum likelihood analysis.

\section{Datasets and methods}
\label{sect2}

\subsection{A Search of $\geq$10~GeV photons}
\label{sect2.1}
From 2008 to 2020, Fermi-LAT has detected more than 200 GRBs. The information about the Fermi-LAT GRBs can be obtained from the second Fermi-LAT GRB catalog~\citep{lat_grb_cat2} and the Fermi website\footnote{\url{https://fermi.gsfc.nasa.gov/ssc/observations/types/grbs/lat_grbs}}. In June 2019, Fermi-LAT team released the Second Fermi-LAT GRB Catalog (the first decade of GRBs detected by the Fermi-LAT), containing 186 LAT GRBs, from a search of a total of 2357 GRBs~\citep{lat_grb_cat2}. In order to enlarge the sample size of possible GRBs with  high-energy photons as much as possible, we also consider GRBs listed in the webpage of Fermi LAT GRBs\footnote{\url{https://fermi.gsfc.nasa.gov/ssc/observations/types/grbs/lat_grbs/table.php}, retrieved on 2019 March 31}. It contains 146 GRBs (including some GRBs in the second Fermi-LAT GRB Catalog). GRB~190114C is the first GRB identified to emit $\sim$TeV emission, so it is also included. In the end, the GRB sample considered in this work contains a total of 199 GRBs.

Fermi-LAT data were downloaded from the {\it Fermi-LAT database\footnote{\url{https://fermi.gsfc.nasa.gov/ssc/data/access/}}},  the ``extended''-class events files were downloaded which include all event classes (including the TRANSIENT class needed for GRB prompt emission analysis) and provide additional information for each event. Fermi Science Tools version 1.2.1 was used to reduce and analyze the 0.1-200 GeV data. For the event class, we chose ``P8R3\_Transient" if the time interval contains the prompt phase, while ``P8R3\_Source"  was selected for the afterglow-only analysis. The zenith angle is constrained to be smaller than 100 degrees to exclude the gamma-ray emission from the Earth albedo.

 In our analysis, the {\it gtselect} tool was used for selection cuts on the event data file. The photon class was selected with 16 for ``transient'' events and 128 for ``source'' events above 100~MeV. The search region has a radius of 10 degrees and is centred on the GRB location. The tool {\it gtmktime} was used to create good time intervals (GTIs) based on selections made using the spacecraft data file variables, and two tools were used to generate exposure maps: {\it gtltcube}, creating a livetime cube, and {\it gtexpmap}, generating an exposure map based on the event selection imposed on the photon file and the livetime cube. To suppress the background, the two diffuse emission components {\it gll\_iem\_v07.fits\footnote{\url{https://fermi.gsfc.nasa.gov/ssc/data/analysis/software/aux/4fgl/gll_iem_v07.fits}}}~(Galactic diffuse emission) and {\it iso\_P8R3\_SOURCE\_V2\_v1.txt\footnote{\url{https://fermi.gsfc.nasa.gov/ssc/data/analysis/software/aux/iso_P8R3_SOURCE_V3_v1.txt}}}~(isotropic diffuse component) were used in our analysis, and sources in the Fermi catalog (4FGL) were included as background sources. Finally, we performed the likelihood fit using the {\it gtlike} tool. All spectra were analyzed as such.

Generally, the best-fit spectral model of LAT afterglow spectra is a power law~\citep[see, e.g.,][]{lat_grb_cat2}:
\begin{equation}
\frac{dN}{dE} = N_0 \left(\frac{E}{E_0}\right)^{-\Gamma},
\end{equation}
where $N_0$ is the normalisation factor, $E_0$ is the pivot photon energy, and $\Gamma$ is the photon index.
The significance of a source can be measured by the test-statistic (TS) value~\citep{lat_grb_cat2}, which is defined as:
\begin{equation}
TS = -2~log \frac{\mathcal{L}_\mathrm{null}}{\mathcal{L}_\mathrm{source}},
\end{equation} where %$\mathcal{L}$ is the Poisson likelihood of observing the given data assuming the best-fit model,
$\mathcal{L}_\mathrm{null}$ is the maximum of the likelihood function for the null hypothesis, and $\mathcal{L}_\mathrm{source}$ is the maximum of the likelihood function for the alternative hypothesis. The detection significance of a source is around $\sim(TS)^{1/2}$ standard deviation (or $\sigma$). The goodness of the different models can be estimated through $\Delta TS$, which is defined as: $\Delta$ $TS_{model1-model2} = TS_{model1}-TS_{model2} = (-2 \mathrm{log}(\mathcal{L}_{null}$ $/$ $\mathcal{L}_{model1}$)) $-$ ($-2\mathrm{log}(\mathcal{L}_{null}$ $/$ $\mathcal{L}_{model2}$)) $=$ $2\mathrm{log}(\mathcal{L}_{model1}$ $/$ $\mathcal{L}_{model2}$), and the significance of improvements is calculated by $\sim(\Delta TS)^{1/2}$ standard deviation (or $\sigma$).

To search for $\geq$10~GeV photons, we analyzed all 199 GRBs with the time interval from the burst onset to 1~day thereafter with the steps described above\footnote{we do not restrict ourselves to the time frame $\geq\,T_\mathrm{90}\times$2 to archive a complete search of $\geq$10~GeV photons from LAT GRBs.}. Following that, we used the {\it gtsrcprob}\footnote{this tool assigns a probability of each photon event that it belongs to a particular source, c.f., \url{https://raw.githubusercontent.com/fermi-lat/fermitools-fhelp/master/gtsrcprob.txt}} tool to estimate the probability of each photon coming from a GRB, $P_\mathrm{source}$. We check all the 199 GRBs to look for $\geq$10~GeV photons associated with a GRB, which are those with $P_\mathrm{GRB}\geq97\%$. Finally, we obtain 67 photons above 10 GeV from 34 GRBs. The properties of these 67 photons and some properties of the 34 GRBs are given in Tables~1--3.

\begin{figure*}
\centering
\includegraphics[width=190mm,height=100mm]{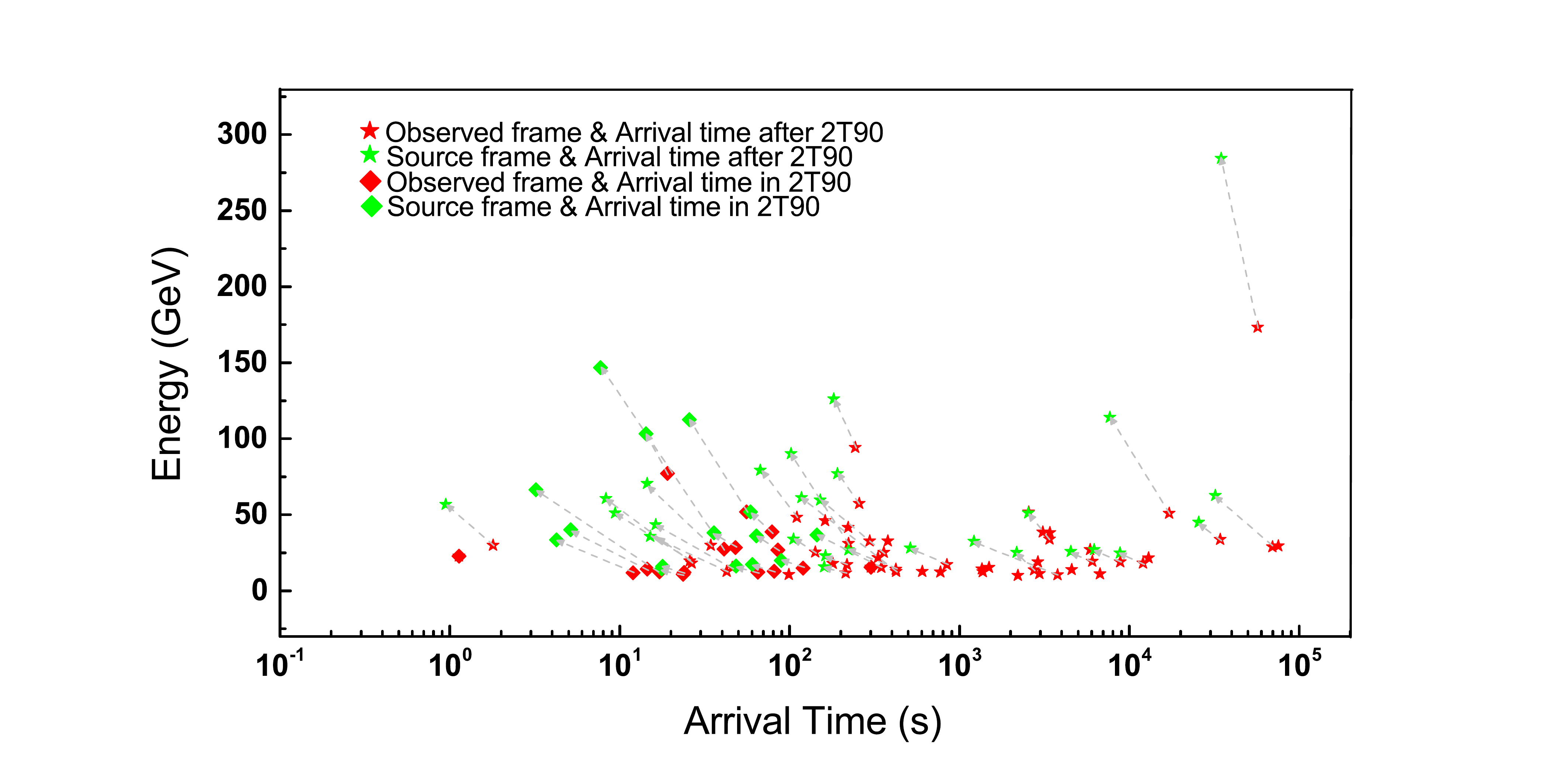}
\caption{The energy and arrival time of photons with detected energy $\geq$10~GeV from Fermi-LAT GRBs from burst onset to 1~day thereafter. Red symbols represent photons arrived within and after 2$\times\,T_\mathrm{90}$, shown in $\blacklozenge$ and $\star$, respectively. Photon energy and arrival time in the source frame are also shown (in green) if measured redshift are available. Dashed arrows connect the same photons in the source frame and the observer's frame.
}
\label{fig1}
\end{figure*}

These 34 GRBs are potential GRBs associated with very high-energy photons. Our samples include the 29 GRBs discovered by Fermi-LAT collaboration~\citep{lat_grb_cat2} with $>$10 GeV photons detected, so our results are consistent with the analysis of the Fermi-LAT collaboration~\citep{lat_grb_cat2}. Among the 34 GRBs, only one GRB (GRB~090510A) is short-duration, and the others are long GRBs. According to when the $>$ 10 GeV photons arrived relative to the GRB trigger time, three categories can be seen:
\begin{itemize}
\item for 27 GRBs, the $\geq$10~GeV photons arrived after $T_\mathrm{90}$ ($T_\mathrm{90}$ is the time duration when 90\% of the prompt emission was detected, see Table~1);
\item for GRBs~160625B, 140619B, 110903A, 080916C, their $\geq$10~GeV photons arrived in $T_\mathrm{90}$ (see Table~2);
\item for GRBs~160509A, 130427A, 090902B, $\geq$10~GeV photons arrived in $T_\mathrm{90}$ and after $T_\mathrm{90}$ (see Table~3).
\end{itemize}
All photons have energy below 100 GeV (in the observer's frame), except for an interesting event -- a 173 GeV photon arrived at 57071~s after trigger with about 97$\%$ probability being associated with GRB~131231A. The energy and arrival time of these photons in observer's and source frames (when a measured redshift is available) are displayed in Figure~\ref{fig1}.

It is clear that most $\geq$10~GeV photons arrived in the afterglow phase, except for GRBs~160625B, 140619B, 110903A, 080916C. It is rather non-trivial to explain late time $\geq$10~GeV photons by synchrotron emission from external shock, as the maximum (cutoff) photon energy radiated by synchrotron is $\sim$ 5 GeV~\citep{PiranNakar10,Abdo2009,Duran2011}, while an additional component in afterglow phase could explain $\geq$10~GeV photons. It is therefore desirable to see if an additional high-energy spectral component can exist in the Fermi-LAT spectra, especially at the high energy end above 10~GeV. We now turn to the analysis of the 30 Fermi-LAT afterglow spectra.

\subsection{Spectral analysis of significantly LAT-detected GRB afterglows with $\geq$10~GeV photons}
\label{sect2.2}

To systematically analyze these 30 GRB afterglows, we adopt a unified end time (86.4~ks post-burst). The analyze time intervals start at 2$\times\,T_\mathrm{90}$ and end at $\sim$86.4~ks for every GRB, where 2$\times\,T_\mathrm{90}$ is adopted to minimise the contamination of the prompt emission at GeV energies. The reasons to derive time-integrated spectra instead of time-resolved ones are two-fold: (1) the number of photons for many of the LAT GRBs especially in the afterglow phase is small; (2) there is no strong indication for change of spectral index during the afterglow phase~\citep{lat_grb_cat}.

Out of the 30 GRBs with $\geq$10~GeV photons arrived after $T_\mathrm{90}$, 25 GRBs are tentatively or significantly detected after 2$\times\,T_\mathrm{90}$ with TS $\ge$4, and this sample will form a basis to study the GeV afterglow spectra as a population.

\begin{figure*}
 \centering
\includegraphics[width=150mm,height=130mm]{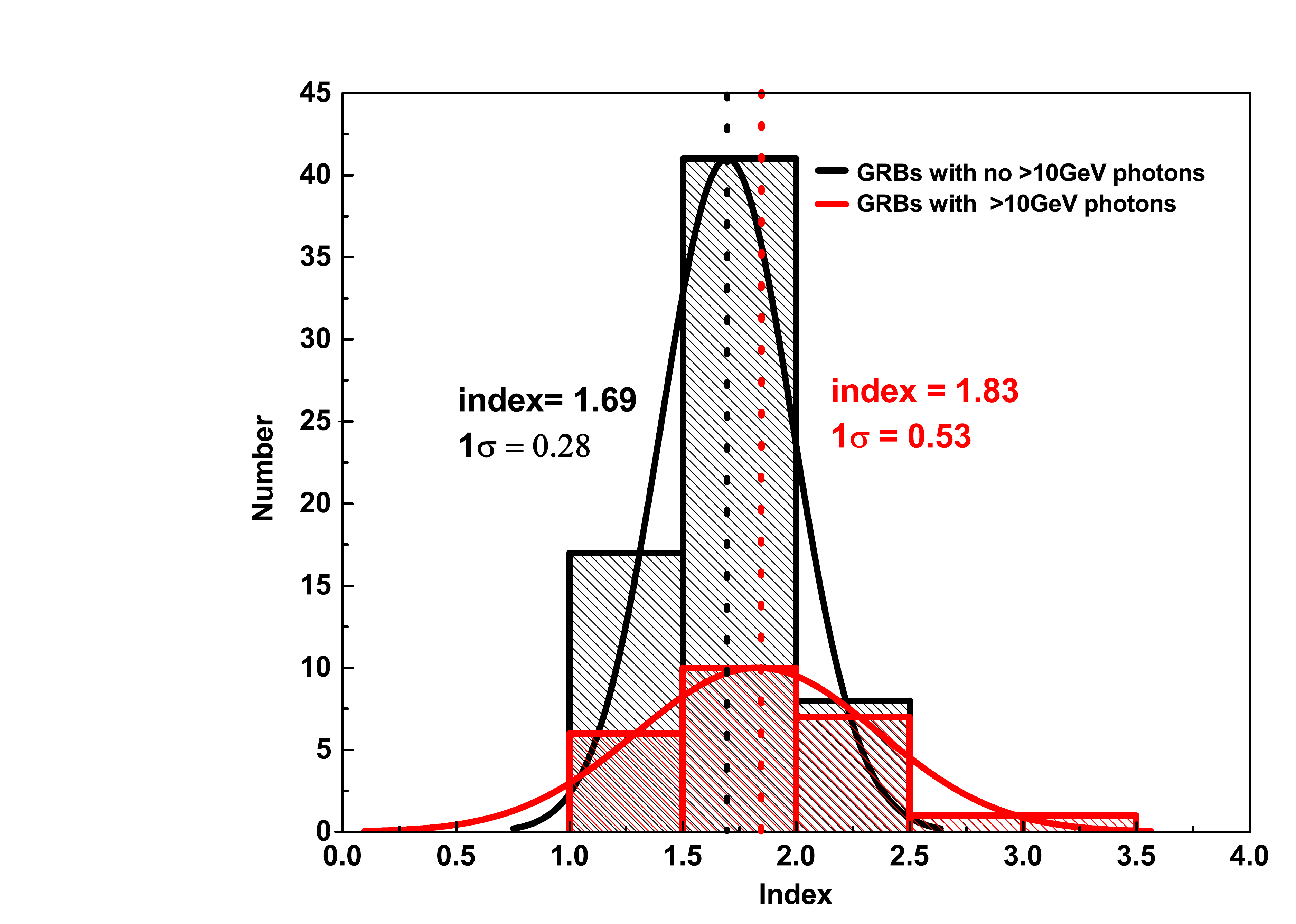}
\caption{The distribution of spectral index from 174 typical GRB afterglows (black) and the 25 GRBs with $\geq$10~GeV photons detected (red) respectively. The spectra of afterglows from 2$\times\,T_\mathrm{90}$ to 1~day are fitted with the PL model in 0.1-200 GeV.}
\label{distribution}
\end{figure*}

We first use the power-law (PL) model to describe the observed spectra. To compare the photon spectral indices between the 25 GRBs with the other 174 LAT GRBs, we plot the index distribution in Figure~\ref{distribution}. The average spectral index of these 25 GRBs ($\Gamma=1.83\pm0.53$) is comparable to that of the typical Fermi-LAT GRBs ($\Gamma=1.69\pm0.28$). We found that the power-law index distribution for the two populations is indeed similar, which indicates that there may not be intrinsic difference between the GeV spectra of the two groups.

To search for an additional higher energy component in these 25 Fermi-LAT GRB afterglows,  We utilize the broken power-law (BPL) model to describe  these GRBs, following \citet{Tam2013} and \citet{Panaitescu2017}. BPL model may improve the fitting if an additional hard component exists in the spectrum. Subsequently, we consider the BPL model for the 25 GRBs:
 \begin{equation}
\frac{dN}{dE} = N_\mathrm{0} \left\{
\begin{array}{ll}
	\left(\frac{E}{E_\mathrm{b}}\right)^{-\alpha}\,\mathrm{if}\,E<E_\mathrm{b} \, , \\
	\left(\frac{E}{E_\mathrm{b}}\right)^{-\beta}\,\mathrm{if}\,E\geq E_\mathrm{b} \, , 
\end{array}
\right.
\end{equation}
where $E_\mathrm{b} $ is the break energy.

\subsection{The possible additional higher energy components in the afterglow spectra }
\label{sect2.3}

To identify the most probable GRBs in which an additional higher energy component may exist in the afterglow. We compare the afterglow spectra using power-law and broken power-law (BPL) representations. There are 18 GRBs that can be well fit by the PL and BPL models. In Figures~3 and~4, we plot the SEDs including the butterfly plots and spectral points. The PL and BPL parameters of the 18 GRBs are shown in Tables~4 and~5. In the cases of GRB~100213C, GRB~101014A, GRB~120916A, GRB~130502B, GRB~160422A, GRB~160521B, and GRB~160509A, small values of the break energy, $E_{break}$, do not allow the low-energy index (i.e., $\alpha$) to be well constrained in the BPL fits.

To search for candidates for which a BPL fits better than a PL, we impose a criterion on the 0.1--200~GeV GRB spectra: based on the butterfly plots, the BPL model should have improvement over the PL model with $\Delta TS_{BPL-PL} >3$, and in the BPL model, the high-energy index ($\beta$) is harder than the low-energy index ($\alpha$), and the spectral points SEDs also show the same behavior that a possible upturn exists in the higher energy. Three categories are drawn:
\begin{itemize}
\item for GRB 131231A, $\Delta TS_{BPL-PL}$ is 13.63 ($\sim$3.7$\sigma$), and an upturn exists above the E$_{break}$. Therefore the BPL model is significantly preferred over the PL model;
\item for GRB~190114C, 171210A, 150902A, 130907A, 130427A, and 090902B, they all satisfy $\Delta TS_{BPL-PL}>$3 and show a possible upturn above the E$_{break}$, therefore the improvements should be tentative or marginal;
\item for the GRBs with $\Delta TS_{BPL-PL}$ < 3, there is no preference for any model.
\end{itemize}
We also plot the spectra of the 7 GRBs ($\Delta TS_{BPL-PL}>$3) in Figure~\ref{combined}, and the break energy $E_\mathrm{b}$ is around 1 GeV for many of these spectra.

For the above 7 GRBs with $\Delta TS_{BPL-PL} >3$, if the additional high energy component exists, it could originate from SSC. In this case, the index ($\beta$) of the high energy component should be harder than the index ($\alpha$) of synchrotron emission at lower energies, the spectrum can be well fit by a BPL model with ${\alpha-\beta=0.5}$~\citep[e.g.,][]{Panaitescu2017}. Therefore, we define BPL$_{\alpha-\beta=0.5}$ model with $\alpha-\beta$ fixed at 0.5 to describe the SSC spectrum.  We find all 7 GRBs could be fit by BPL$_{\alpha-\beta=0.5}$ model.  It supports that an SSC-like spectrum is acceptable for these 7 cases. The fit results are presented in Table~4.

To summarise, we found that the BPL improvement for GRB 131231A is significant, with a maximum $\Delta TS_{BPL-PL}$ = 13.63 ($\sim$3.7$\sigma$) and an upturn above E$_{break}=1.6\pm0.8$~GeV.  For GRB~190114C, GRB 171210A, GRB 150902A, GRB 130907A, GRB 130427A, and GRB 090902B, we regard the improvement of the BPL fit over PL to be tentative or marginal, since these 6 GRBs show $\Delta TS_{BPL-PL}>3$ and a possible upturn above E$_{break}$. The spectra of the other GRBs can be fitted equally well by the two models.

\subsection{The possible additional higher energy components in the prompt spectra}
\label{sect2.4}

For these 7 GRBs, it is desirable to see whether the additional high energy component exists in prompt emission as well. For that, we check whether the spectrum exhibits an upturn during the prompt phase or shortly after. Therefore, we ran the unbinned analysis chain in $T_\mathrm{0}$ to $T_\mathrm{0}+$2$\times\,T_\mathrm{90}$ (where $T_\mathrm{0}$ is the GRB onset time) to get a power-per-decade spectra $F$, defined as \begin{equation}
F = E \frac{dN}{dE}  \propto \left(E\right)^{-\gamma}.
\end{equation} Only GRB~090902B, GRB~130427A and GRB~190114C show significant prompt emission. For these 3 GRBs, the high enough number of photons allow us to plot bin-wise spectra.
The power-per-decade spectra $F$ of these three GRBs are shown in Figure~\ref{sample4}. We use the PL and BPL models to fit the data points with linear regression, and the $\chi^2/d.o.f.$ values show that the BPL model gives a better fit for GRB~130427A than the PL model does. It is also apparent that an upturn exists in GRB 130427A during the prompt phase (i.e., $T_\mathrm{0}$ to $T_\mathrm{0}$ $+$ 2$\times\,T_\mathrm{90}$), and the break energy is $\sim$2~GeV. In the prompt phase of GRB 190114C, the time-resolved spectrum shows that the spectrum evolves from a BPL to a PL over the course of the prompt phase~\citep{Ajello2020, Vikas2020}. In our analysis, we used time-integrated spectra over the whole prompt phase, and both models can describe the spectrum equally well.

\section{Discussion}

We found a group of GRB afterglows where an additional high energy component above $\sim$1~GeV may exist (see Figure~\ref{sample3}). By comparing the TS values of these 7 GRBs assuming either the PL or the BPL model, we found that the BPL model is the preferred model for GRB~131231A, as in this case $\Delta TS_{BPL-PL} =$ 13.63 ($\sim$3.7$\sigma$) is found, supported by an upturn above the E$_{break}$. We found tentative (or marginal) evidence that the BPL model could be superior to the PL model (3 $<\Delta TS_{BPL-PL}<$ 9 and a possible upturn above E$_{break}$) for the cases GRB~190114C, GRB 171210A, GRB 150902A, GRB 130907A, GRB 130427A, and GRB 090902B. One should note that, as time evolves in the afterglow phase, the break energy could vary or even move outside of the instrument energy range, smoothing any sharp upturn when summing over spectra at different times. Therefore, evidence of such an upturn in {\it time-integrated} afterglow spectra could strengthen the existence of an upturn.

%%%For GRB130427A and GRB190114C, the additional spectral component appears in prompt and afterglow phase.

It is hard to explain the late-time $\geq$10~GeV photons by synchrotron emission (which is typically used to explain afterglow emission below $\sim$1~GeV), and the evidence of an upturn above a break energy of $\sim$1~GeV requires an additional spectral component at high energy. A leading explanation for this additional component is the SSC radiation in the external forward shock~\citep[see, e.g.,][for the case of GRB~190114C]{MagicB}. It has been predicted over the past two decades~\citep{Panaitescu1998,Wei1998,Sari2001,Zhang2001,Meszaros2004,
Fan2008,Galli2008,Nakar2009,PiranNakar10,Lemoine2015} and the VHE afterglow emission detected in GRB~180720B and GRB~190114C provides a strong evidence for such an SSC component origin~\citep{MagicB,HESS}.

If the additional component is caused by SSC, in the synchrotron-SSC scenario, the Fermi-LAT spectrum below $E_\mathrm{b} $ representing the synchrotron emission would have a photon index larger than 2, while the SSC emission starts to dominate above a few GeV making the spectra ``anomalously" turn up above the break energy. This is broadly consistent with the 7 GRBs with an upturn at $\sim$1~GeV.

In order to investigate whether the SSC emission can explain the new component, we fit the observations of GRB~131231A whose spectrum upturn is the most significant in terms of $\Delta TS_{BPL-PL}$ with a simple SSC model presented by ~\citet{Sari1998} and \citet{Sari2001}. This model gives an analytic approximation of synchrotron and SSC spectra ignoring the Klein-Nishina effects. This approximation is adequate given the limited photon statistics up to 200~GeV. The result of modeling to the Swift XRT and Fermi-LAT data of GRB~131231A, within the framework of the theory of afterglow emission from external forward shocks~\citep{Sari1998,Sari2001}, is shown in Figure~\ref{modeling}. The \texttt{Swift}-XRT data are analyzed by HEASOFTv6.25.
The X-ray spectra are fitted with an absorbed, redshifted power-law model considering the column density of the Galactic hydrogen of $n_\mathrm{H}=0.028\times 10^{22}$~cm$^{-2}$, that of the intrinsic hydrogen of $n_\mathrm{H}=0.24\times 10^{22}$~cm$^{-2}$, and a redshift of $z=0.642$\footnote{\url{https://www.swift.ac.uk/xrt_spectra/00020336/}}. The X-ray spectral index is obtained as $\Gamma = 2.76_{-1.61}^{+1.52}$, which gives an unabsorbed 0.3--10.0\,keV flux of $(2.09 \pm 0.9) \times 10^{-13}$~erg\,cm$^{-2}$\,s$^{-1}$. 

We found that the SSC model can explain the upturn of the Fermi-LAT observed spectrum if the conditions at the source are the following. The energy of the spherical shock is roughly estimated to be $E\approx9.2\times10^{52}$\,erg through the observed energy fluence ($\approx$3.7$\times 10^{-4}$\,erg\,cm$^{-2}$) from 100\,MeV to 200\,GeV and 60\,s to 1\,day after the burst onset. The relativistic shock propagates into a constant surrounding density n$\thickapprox$1\,cm$^{-3}$, accelerating the electrons and forming a power law distribution of Lorentz factor $\gamma_{\rm e}$ with an index \boldsymbol{$p=2.13_{-0.05}^{+0.25}$}. A constant fraction \boldsymbol{$\epsilon_{\rm e}=0.16_{-0.03}^{+0.04}$} of the shock energy goes into the electrons and about 20\% of the electron energy is radiated away via the synchrotron and SSC emission. A fraction \boldsymbol{$\epsilon_{\rm B}=1_{-0.5}^{+39}\times10^{-5}$} of the shock energy goes into amplifying the magnetic fields behind the shock. These three (n is fixed) free-parameter values of $p$, $\epsilon_{\rm e}$ and $\epsilon_{\rm B}$ are resulted from an acceptable fit with the best goodness of fit $\chi^{2}/\rm d.o.f$=4.23/(8-3) and their ranges correspond to one $\sigma$ confidence interval.

Corresponding to these values of fitting parameters and the observed time, the emitting relativistic electrons are in the regime of slow cooling. We note that $\epsilon_e\gg\epsilon_B$, thus the necessary condition for efficient production of SSC radiation can be satisfied~\citep{Sari1998,Sari2001}. These results are consistent with a previous work by~\citet{Liu2014} on the same GRB, who focus especially on the origin of the GeV afterglow before 1000s post-burst.

Based on the spectra shown in Figure~\ref{combined}, the spectra at high energies are harder than low energies, if both components originate from the same shock, the low energy component is synchrotron emission and higher energy components possibly are SSC emission. In this case, the synchrotron emission should be produced above the cooling break, and $\alpha = p/2$, SSC spectral index in the Fermi-LAT band should be $\beta= (p-1)/2$, such that $(\alpha - \beta) = 1/2$~\citep{Panaitescu2017}. In our BPL results, we indeed found that $(\alpha - \beta) \sim0.52$ for GRB~130427A and $(\alpha - \beta) \sim 0.49$ GRB~090902B, satisfying $(\alpha - \beta) = 1/2$. We also found that the BPL model with $\alpha-\beta=0.5$ being fixed gives acceptable fits for the 7 GRBs.

We compared the spectra of these 7 GRBs above 20~GeV with the sensitivity of Cherenkov Telescope Array(CTA) and Major Atmospheric Gamma Imaging Cherenkov telescope(MAGIC)~\citep{Giuseppe2019}. We assume z=1 for GRB~171210A and GRB~150902A. After considering the EBL absorption~\citep{Franceschini2017} and performing an optimistic and extreme assumption that the spectrum extends with the same high-energy index ($\beta$) up to 10~TeV\footnote{this assumption is an extreme case, ignoring the cutoff in the natural SSC spectra slope, the Klein-Nishina effect and the pair opacity in the very high energy}, the result is shown in Figure~\ref{sensitivity}. It can be seen that given favourable observing conditions, GRB~130427A, GRB~131231A, and GRB~190114C can likely be detected by CTA and MAGIC. The question on whether the additional component commonly exists in Fermi-LAT GRBs can be ultimately tested by a large number of GRB observations using the LHAASO-WCDA detector~\citep{lhaaso}, CTA array~\citep{CTA} and the currently operating Cherenkov telescopes.

The photon index distribution of the 25 GRBs (with $\geq$10~GeV detected) is similar to other 174 Fermi-LAT GRBs, and both are indeed rather hard ($\Gamma<$2) at the LAT band (see Figure~\ref{distribution}). This result lends support to the idea that these 25 GRBs do not form a distinctive class of their own, and their LAT afterglow spectra are not dissimilar to most other Fermi-LAT GRBs.

%\textcolor{red}{This possibility can be tested with more observations. If this is true, the reason of non-detection for the very high-energy ($\geq$100~GeV) photons of some GRBs, may be due to Klein-Nishina cut-off~\citep{Nakar2009,aliu14} , internal $\gamma\gamma$ absorption\citep{Panaitescu2017,Derishev2019}, or attenuation by the extragalactic background light~\citep{MagicA}.

\section{Conclusions}

Based on the analysis of a large sample of Fermi-LAT afterglows, there is no conclusive evidence that an additional higher energy component commonly exists in Fermi-LAT GRB afterglows. On one hand, a group of GRBs, namely, GRB~190114C, GRB 171210A, GRB 150902A, GRB 131231A, GRB 130907A, GRB 130427A, and GRB 090902B, marginaly or significantly supported by the plausible existence of an upturn above the break energy in the more preferred BPL model, might require such a component at the high energy end. For GRB~130427A, the spectral upturn appears in both prompt and afterglow phases. On the other hand, the BPL model is not required in the large majority of cases, as there is no improvement of the BPL fit compared to the PL fit for all other cases.

If the additional component really exists, it may be resulted from an SSC component. Besides, we find that the SSC model can explain the upturn in GRB~131231A under acceptable typical parameter values of GRBs. Another evidence for the possible SSC component is provided by GRB~130427A and GRB~090902B  whose $\alpha - \beta$ ($\sim$0.52 and $\sim$0.49, respectively) largely satisfy $\alpha - \beta = 0.5$, and all 7 GRBs could be satisfactorily fit by the BPL$_{\alpha-\beta=0.5}$ model. These suggest that SSC emission exists in a group of GRBs (regardless of whether they are detected by ground-based VHE telescopes), and the SSC emission may last from the prompt to afterglow phases, as seen in GRB~130427A.

The possible additional high-energy component found in a group of Fermi-LAT GRBs is clearly in contradiction with a single component such as synchrotron emission for these afterglows, and SSC emission is a widely discussed mechanism to explain the additional high-energy component. Current and future VHE observations using CTA, LHAASO, and currently operating Cherenkov telescopes will provide important constraints on whether SSC emission commonly exists among GRB afterglows.

\section*{Acknowledgments}
This work is supported by the National Natural Science Foundation of China (NSFC) grants 11633007 and U1731136, and Guangdong Major Project of Basic and Applied Basic Research (Grant No. 2019B030302001). Y.Z. thanks NSFC grant number 11973099 for financial support. This work made use of the LAT data and science tools available at the Fermi Science Support Center.

\label{sect3}

\begin{table*}
\centering
\label{table1}
\caption {Highest energy events of \textit{Fermi}-LAT GRBs for which $\geq$10~GeV photons arrive after $T_\mathrm{90}$ only \label{tab_energymax_GRB}}
\begin{tabular}{ccccccccc}
\hline \hline
 \multicolumn{4}{c}{GRB Properties} & \multicolumn{5}{c}{Photon Properties} \\ \cline{1-4}\cline{5-9}
Name &.  $T_\mathrm{90}$\tablefootmark{a}$_{of}$   &   $T_\mathrm{90}$\tablefootmark{b}$_{sf}$    &  Redshift\tablefootmark{c}   &  Energy\tablefootmark{d}$_{of}$    &    Arrival time\tablefootmark{e}$_{of}$      &   Energy\tablefootmark{f}$_{sf}$    &    Arrival Time\tablefootmark{g}$_{sf}$    &   P\tablefootmark{h}   \\
\hline
190114C & 116.0 & 81.4 & 0.43 &18.94& 8849.2 & 27.0 & 6210.0 &1 \\
171210A & 143.1 & $-$ & $-$  &12.49& 1374.6& $-$ & $-$ & 1 \\
171022A &13.3 & $-$ & $-$ &13.98& 4582.8 & $-$ & $-$ & 1\\
171010A & 107.3 & 80.7& 0.33&19.00& 2891.0 & 25.3 & 2173.7 & 1 \\
160623A & 107.8 & 78.9 &0.37 & 18.21 & 12039.8  &24.9& 8807.5 & 1 \\
160521B & 2.8 & $-$ & $-$ &12.69&423.7  &$-$ & $-$ &1 \\
 160422A & 12.3 & $-$ & $-$ &12.29 &771.4 & $-$ & $-$ & 1  \\
160310A & 25.6  & $-$ & $-$ &26.99 &5886.0 & $-$ & $-$  & 1  \\
150902A & 13.6 & $-$ & $-$ &10.59 &98.9 & $-$ & $-$ & 1 \\
140928A & 17.9& $-$  & $-$ &51.76&2555.2 & $-$ & $-$ & 1 \\
 & $-$  & $-$ & $-$ &38.62 &3100.9 &$-$&$-$ & 1 \\
140810A &81.7 & $-$ & $-$ &15.38 &1490.3 & $-$ & $-$& 1  \\
140416A &31.7 & $-$ & $-$ &10.08 &2208.3& $-$ & $-$ & 1  \\
140206B &27.3  & $-$ & $-$ &10.96  &6736.7 & $-$ & $-$& 1 \\
 & $-$  & $-$ &$-$ &29.39 &75494.0&$-$&$-$ & 1   \\
131231A&31.2 &19.0 &0.64 &48.29 &110.4 & 79.3 &67.2 & 1 \\
 & $-$  & $-$  & $-$ &17.14 &844.0 & 28.1 &514.0 & 1\\
 & $-$  & $-$ & $-$&173.17 &57071.0 &284.4 &34757.0  &0.97  \\
130907A&115.0 &51.4  &1.24 &50.96 &17161.0 &114.1 &7668.0 & 1\\
130502B&24.3 & $-$ & $-$ &31.10&222.1 & $-$ & $-$ & 1 \\
120919B&118.0 & $-$ & $-$ &12.70 &605.0 & $-$ & $-$ & 1 \\
120916A&53.0 & $-$ & $-$ &21.57 &13004.8 & $-$ & $-$ & 1 \\
120526A&43.6  & $-$ & $-$ &14.30 &1354.3 & $-$ &$-$ & 1\\
101014A&449.4  & $-$ & $-$ &13.60 &2750.7& $-$ & $-$ & 1 \\
 &$-$& $-$ & $-$ &11.20 &2962.0 & $-$ & $-$ & 1 \\
100511A &42.4 & $-$ & $-$ &46.00 &161.9& $-$ & $-$ & 1\\
 &$-$& $-$ & $-$ &18.00 &179.8 & $-$ & $-$ & 1 \\
100414A&26.5 &11.1 &1.37 &29.80 &34.4 &70.6&14.5 & 1\\
 &$-$& $-$ & $-$ &25.13 &359.5 &59.5 &151.8 &1 \\
100213C&60.0  & $-$ & $-$ &34.00 &3389.0 & $-$ & $-$ & 1 \\
100116A&102.5 & $-$ & $-$ &32.64 &379.2 & $-$ & $-$ & 1 \\
 &$-$& $-$ & $-$ &13.35 &296.7 & $-$ & $-$ & 1 \\
090926A&20.0 &6.4  &2.11 &19.46 &25.8 &60.5 &8.3 & 1 \\
 &$-$& $-$ &$-$&10.42 &3786.0 &32.4 &1217.4 & 1 \\
090510A&1.0 &0.5  &0.90&29.91 &1.8 &56.8 &0.9 & 1\\
090427A&12.3 & $-$ & $-$ &14.10 &422.9 & $-$ & $-$ & 1 \\
\hline
\end{tabular}
\tablefoot{The unit of energy is GeV, $T_\mathrm{90}$ and arrival times are in seconds.\\
\tablefoottext{a,d,e} { Observation Frame$_{of}$; }
\tablefoottext{b,f,g}{ Source Frame$_{sf}$; }
\tablefoottext{a$^{ref}$}{ https://heasarc.gsfc.nasa.gov/cgi-bin/W3Browse/w3hdprods.pl; }
\tablefoottext{c$^{ref}$}{ http://www.mpe.mpg.de/~jcg/grbgen.html; }
\tablefoottext{h}{ Probability of the photon being associated with that GRB. }
}
\end{table*}

\begin{table*}
\centering
\label{table2}
\caption {Events with the highest energy of \textit{Fermi}-LAT GRBs for which $\geq$10~GeV photons arrive in $T_\mathrm{90}$ only\label{tab_energymax_GRB}}
\begin{tabular}{ccccccccc}
\hline \hline
 \multicolumn{4}{c}{GRB Properties} & \multicolumn{5}{c}{Photon Properties} \\ \cline{1-4}\cline{5-9}
Name &.  $T_\mathrm{90}$\tablefootmark{a}$_{of}$   &   $T_\mathrm{90}$\tablefootmark{b}$_{sf}$    &  Redshift\tablefootmark{c}   &  Energy\tablefootmark{d}$_{of}$    &    Arrival time\tablefootmark{e}$_{of}$      &   Energy\tablefootmark{f}$_{sf}$    &    Arrival Time\tablefootmark{g}$_{sf}$    &   P\tablefootmark{h}   \\
\hline
160625B & 453.4 & 188.4  & 1.41&15.30& 347.5 & 36.8  &144.4 & 1 \\
140619B&2.8& $-$ & $-$ &22.74&1.1 & $-$ & $-$ & 1 \\
110903A&341.3 & $-$ & $-$ &15.60 &301.0 & $-$ & $-$ & 1 \\
080916C&63.0 &11.8  &4.35 &12.42 &17.2 &66.5 &3.2 & 1\\
&$-$& $-$ &$-$&27.43 &41.1&146.7 &7.7  & 1 \\
\hline
\end{tabular}
\tablefoot{The unit of energy is GeV, $T_\mathrm{90}$ and arrival times are in seconds;\\
\tablefoottext{a,d,e} { Observation Frame$_{of}$; }
\tablefoottext{b,f,g}{ Source Frame$_{sf}$; }
\tablefoottext{a$^{ref}$}{ https://heasarc.gsfc.nasa.gov/cgi-bin/W3Browse/w3hdprods.pl;}
\tablefoottext{c$^{ref}$}{ http://www.mpe.mpg.de/~jcg/grbgen.html;}
\tablefoottext{h}{ Probability of the photon being associated with that GRB. }
}
\end{table*}

\begin{table*}
\centering
\label{table2}
\caption {Events with the highest energy of \textit{Fermi}-LAT GRBs  for which $\geq$10~GeV photons arrive both within $T_\mathrm{90}$ and thereafter \label{tab_energymax_GRB}}
\begin{tabular}{ccccccccc}
\hline \hline
 \multicolumn{4}{c}{GRB Properties} & \multicolumn{5}{c}{Photon Properties} \\ \cline{1-4}\cline{5-9}
Name &.  $T_\mathrm{90}$\tablefootmark{a}$_{of}$   &   $T_\mathrm{90}$\tablefootmark{b}$_{sf}$    &  Redshift\tablefootmark{c}   &  Energy\tablefootmark{d}$_{of}$    &    Arrival time\tablefootmark{e}$_{of}$      &   Energy\tablefootmark{f}$_{sf}$    &    Arrival Time\tablefootmark{g}$_{sf}$    &   P\tablefootmark{h}   \\
\hline
160509A & 369.7 & 170.4   & 1.17  &51.90 & 55.7 & 112.6  & 25.7  & 1 \\
 & $-$  & $-$ &$-$ &41.51&221.4 & 90.1 & 102.0  & 1 \\
 & $-$  & $-$ & $-$ &28.90&69688.6 & 62.7 & 32114.6 & 1   \\
 130427A&138.2  &103.1  &0.34 &94.12 &243.6 &126.1 &181.8& 1\\
 & $-$  & $-$ & $-$ &77.11 &19.1 &103.3 &14.2 & 1  \\
 & $-$  & $-$ & $-$ &57.42 &256.7 &76.9 &191.6 & 1  \\
 & $-$  & $-$ & $-$ &38.67 &78.8 &51.8 &58.8 & 1  \\
 & $-$  & $-$ & $-$ &38.19 &3410.3 &51.2 &2545.2 & 1 \\
 & $-$  & $-$ & $-$ &33.65 &34366.6 &45.1 &25648.6 & 1  \\
 & $-$  & $-$ & $-$ &28.41 &48.0 &38.1 &35.8 & 1  \\
& $-$  & $-$ & $-$  &26.90 &85.2 &36.1 &63.6 & 1  \\
 & $-$  & $-$ & $-$ &25.36 &141.5 &34.0 &105.6 & 1  \\
 & $-$  & $-$ & $-$ &19.27 &6063.0 &25.8 &4524.9 & 1  \\
 & $-$  & $-$ & $-$ &17.08 &217.9 &22.9 &162.6 & 1  \\
 & $-$  & $-$ & $-$ &14.90 &119.7&20.0 &89.4 & 1  \\
 & $-$  & $-$ & $-$ &12.87 &80.9 &17.2 &60.4 & 1  \\
& $-$  & $-$ & $-$  &12.18 &64.9 &16.3 &48.4 & 1  \\
 & $-$  & $-$ & $-$ &12.00 &23.9 &16.1 &17.8 & 1  \\
& $-$  & $-$ & $-$  &11.74 &214.4 &15.7 &160.0 & 1  \\
 & $-$  & $-$ & $-$ &10.85 &23.7 &14.5 &17.6 & 1  \\
090902B&19.3 &6.8  &1.82 &21.72 &332.2 &61.3 &117.7 & 1\\
&$-$& $-$ & $-$  &18.11 &26.5 &51.1 &9.4 & 1 \\
&$-$& $-$ & $-$  &15.40 &45.9 &43.5 &16.3 & 1 \\
&$-$& $-$ & $-$   &14.22 &14.5 &40.1 &5.1 & 1 \\
 &$-$& $-$ & $-$  &12.66 &42.7 &35.7 &15.1& 1 \\
&$-$& $-$ & $-$   &11.89 &12.0 &33.6 &4.2& 1 \\
\hline
\end{tabular}
\tablefoot{The unit of energy is GeV, $T_\mathrm{90}$ and arrival times are in seconds;\\
\tablefoottext{a,d,e} { Observation Frame$_{of}$; }
\tablefoottext{b,f,g}{ Source Frame$_{sf}$; }
\tablefoottext{a$^{ref}$}{ https://heasarc.gsfc.nasa.gov/cgi-bin/W3Browse/w3hdprods.pl;}
\tablefoottext{c$^{ref}$}{ http://www.mpe.mpg.de/~jcg/grbgen.html;}
\tablefoottext{h}{ Probability of the photon being associated with that GRB. }
}
\end{table*}

\begin{table*}
\centering
\label{table3}
\caption {The fitting parameters of the PL and BPL model of 7 GRBs which show a possible upturn in the afterglow spectra.}
\begin{tabular}{c|cc|cccc|c}
\hline \hline
 & \multicolumn{2}{c}{PL model} & \multicolumn{4}{c}{BPL $\&$ BPL$_{\alpha-\beta=0.5}$ model} \\ \cline{1-4}\cline{5-8}
Name   &  $\Gamma$  & TS & $\alpha$    &  $\beta$  &  E$_b$    & TS   &  $\Delta$TS\\
\hline
190114C  & 2.37 $\pm$ 0.41 & 20.19 & 3.71 $\pm$ 0.89 & 1.39 $\pm$ 0.46 & 696 $\pm$ 283 & 27.92 & 7.73 \\
& -- & -- & 2.72 $\pm$ 0.37 & $\alpha$-0.5 & 631 $\pm$ 202 & 22.93 & 2.74\\\hline
171210A  & 2.33 $\pm$ 0.28 & 34.31  & 2.79 $\pm$ 0.49 & 1.48 $\pm$ 0.78 & 1368 $\pm$ 755 & 37.70 & 3.39 \\
& -- & -- & 2.51 $\pm$ 0.31 & $\alpha$-0.5 & 1260 $\pm$ 424 & 36.18 & 1.87\\\hline
150902A  &2.52 $\pm$ 0.31 & 41.29 & 3.41 $\pm$ 0.55 & 1.55 $\pm$ 0.41 & 637 $\pm$ 222 & 49.28 & 7.99\\
& -- & -- & 2.92 $\pm$ 0.45 & $\alpha$-0.5 & 584 $\pm$ 274 & 44.39 & 3.10\\\hline
131231A  & 1.72 $\pm$ 0.11 & 165.31 & 2.03 $\pm$ 0.58 & 1.32 $\pm$ 0.21 & 1586 $\pm$ 696 & 178.94 & 13.63 \\
& -- & -- & 1.93 $\pm$ 0.11 & $\alpha$-0.5 & 1585 $\pm$ 783 & 178.60 & 13.29\\\hline
130907A  & 1.79 $\pm$ 0.30 & 39.27 & 2.54 $\pm$ 0.56 & 1.08 $\pm$ 0.56& 1741 $\pm$ 1007 & 42.89 & 3.62 \\
& -- & -- & 2.05 $\pm$ 0.28 & $\alpha$-0.5 & 1712 $\pm$ 952 & 41.23 & 1.96\\\hline
130427A & 2.22 $\pm$ 0.10 &611.14 & 2.38 $\pm$ 0.14  & 1.86 $\pm$ 0.28 & 1080 $\pm$ 545 & 615.00 & 3.86 \\
& -- & -- & 2.40 $\pm$ 0.15 & $\alpha$-0.5 & 1080 $\pm$ 361 & 614.97 & 3.83\\\hline
090902B  & 1.86 $\pm$ 0.11 & 458.53  & 2.31 $\pm$ 0.16 & 1.82  $\pm$ 0.14  & 664 $\pm$ 390 & 463.19 & 4.66 \\
& -- & -- & 2.40 $\pm$ 0.18 & $\alpha$-0.5 & 584 $\pm$ 314 & 462.91 & 4.38\\\hline
\hline
\end{tabular}
\tablefoot{
The unit of E$_b$ is MeV. All spectral fits are made using 0.1--200~GeV data collected between 2$\times\,T_\mathrm{90}$ and one day after. $\Delta TS$ is $TS_{BPL}-TS_{PL}$ and $TS_{BPL_{\alpha-\beta=0.5}}-TS_{PL}$.
}
\end{table*}

\begin{table*}
\centering
\label{table4}
\caption {The fitting parameters of the PL and BPL model of 11 GRBs which do not show an upturn in the afterglow GeV spectra.}
\begin{tabular}{c|cc|cccc|c}
\hline \hline
 & \multicolumn{2}{c}{PL model} & \multicolumn{4}{c}{BPL model} \\ \cline{1-3}\cline{4-8}
Name   &  $\Gamma$  & TS & $\alpha$    &  $\beta$  &  E$_b$    & TS &  $\Delta$TS   \\
\hline
171022A & 1.18 $\pm$ 0.73 & 13.03 &2.33 $\pm$ 2.05 &1.08 $\pm$ 0.46 & 1166 $\pm$ 623 & 13.35 & 0.32 \\
171010A & 2.07 $\pm$ 0.12 & 162.04 &1.92 $\pm$ 0.24 &2.31 $\pm$ 0.38 & 826 $\pm$ 407 & 164.15 & 2.11 \\
160623A & 1.85 $\pm$ 0.10 & 145.83 &1.78 $\pm$ 0.23 & 1.91 $\pm$ 0.41 & 637 $\pm$ 222 & 145.66 & -0.17 \\
160310A & 1.24 $\pm$ 0.64 & 10.02 &1.81 $\pm$ 1.20 & 1.10 $\pm$ 0.69& 1282 $\pm$ 629 & 10.18 & 0.16 \\
140928A & 1.14 $\pm$ 0.31 & 33.90 & 1.92 $\pm$ 0.71 &0.84 $\pm$ 0.41 & 1999 $\pm$ 81 &36.01 & 1.11 \\
140810A & 1.66 $\pm$ 0.20 & 59.29 &0.01 $\pm$ 0.54 & 1.85  $\pm$ 0.29  & 559 $\pm$ 102 & 60.33 & 1.04 \\
140416A & 1.65 $\pm$ 0.40 & 16.52 &1.38 $\pm$ 0.02 &2.00 $\pm$ 0.64 & 516 $\pm$ 177 & 17.50 & 0.98 \\
140206B & 1.98 $\pm$ 0.11 & 177.00 &2.08 $\pm$ 0.21 &1.89 $\pm$ 0.21 & 917 $\pm$ 755 & 177.88  & 0.88 \\
100414A & 2.02 $\pm$ 0.14 & 98.58 &0.73 $\pm$ 0.65 &2.35 $\pm$ 0.26 & 310 $\pm$ 128 & 103.39 & 4.81 \\
090926A & 1.86 $\pm$ 0.08 & 176.70 &1.87 $\pm$ 0.56 & 1.86 $\pm$ 0.22& 999 $\pm$ 82 & 176.71 & 0.01 \\
090510A & 2.09 $\pm$ 0.07 & 381.12 & 2.00 $\pm$ 0.09 &2.51 $\pm$ 0.36& 2024 $\pm$ 186 & 383.79 & 2.67 \\
\hline
\end{tabular}
\tablefoot{
The unit of E$_b$ is MeV. All spectral fits are made using 0.1--200~GeV data collected between 2$\times\,T_\mathrm{90}$ and one day after. $\Delta TS = TS_{BPL}-TS_{PL}$.
}
\end{table*}
\newpage

\newpage

\begin{figure*}
\includegraphics[width=92mm,height=85mm]{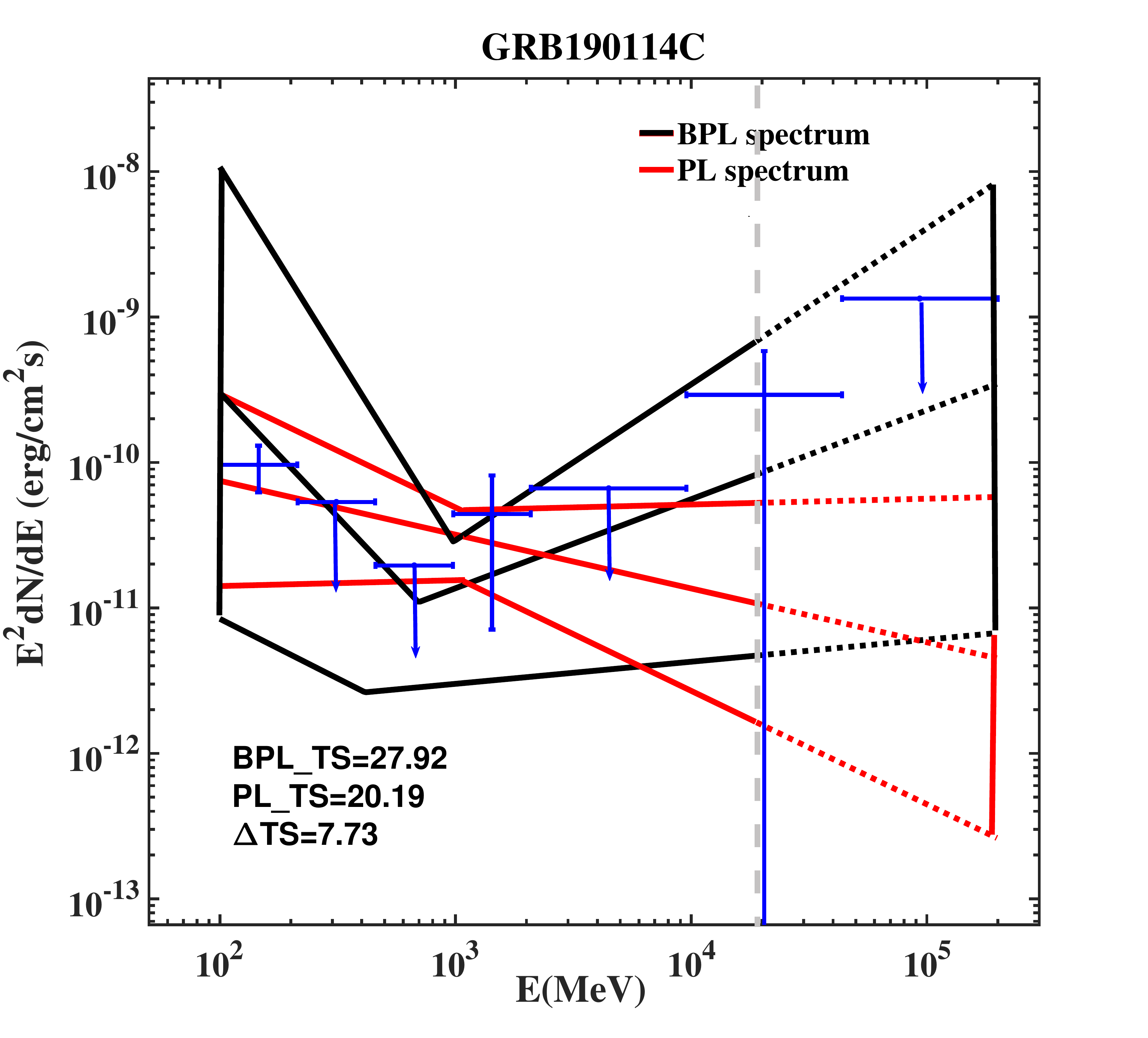}
\includegraphics[width=92mm,height=85mm]{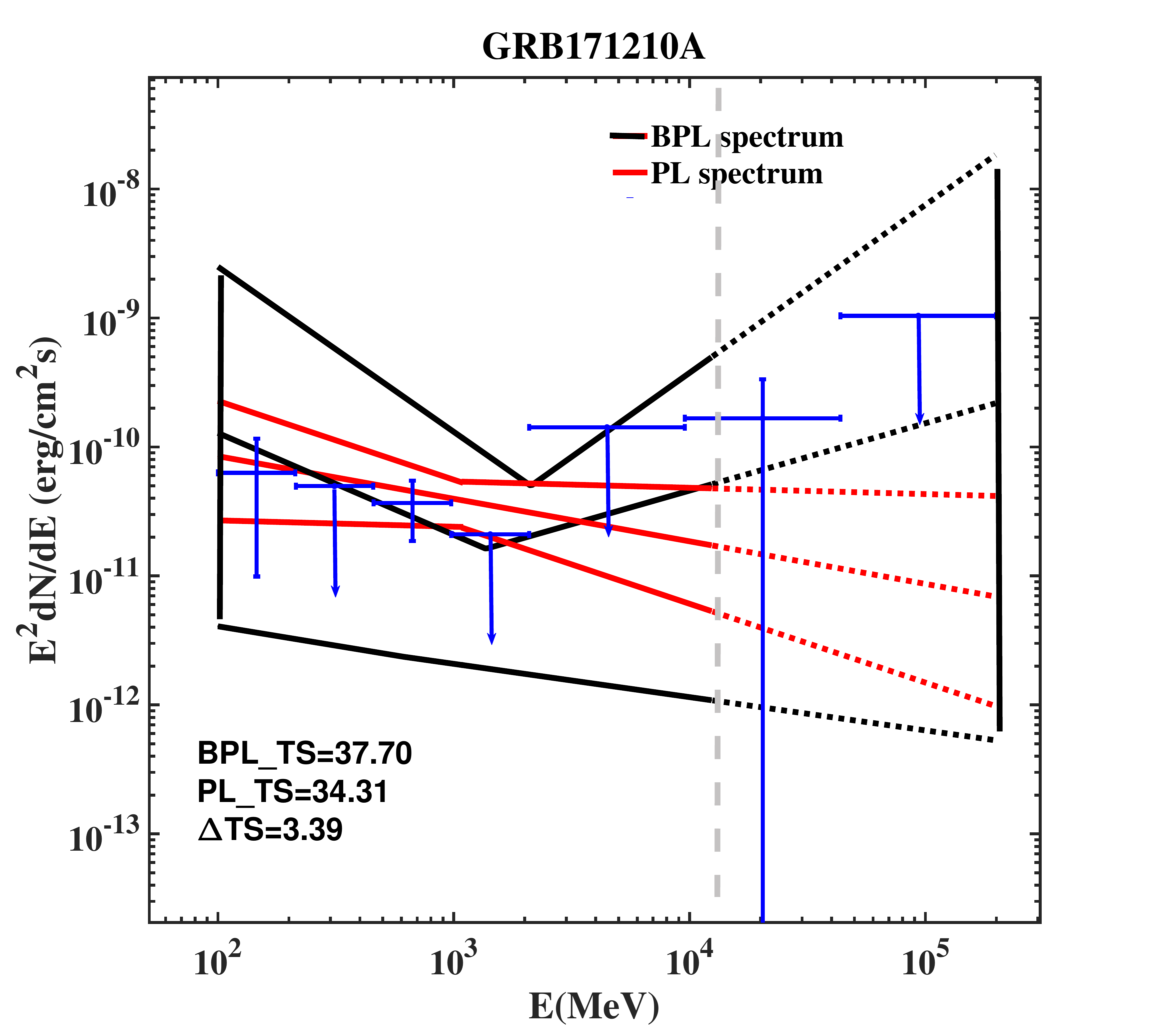}\\\\
\includegraphics[width=92mm,height=85mm]{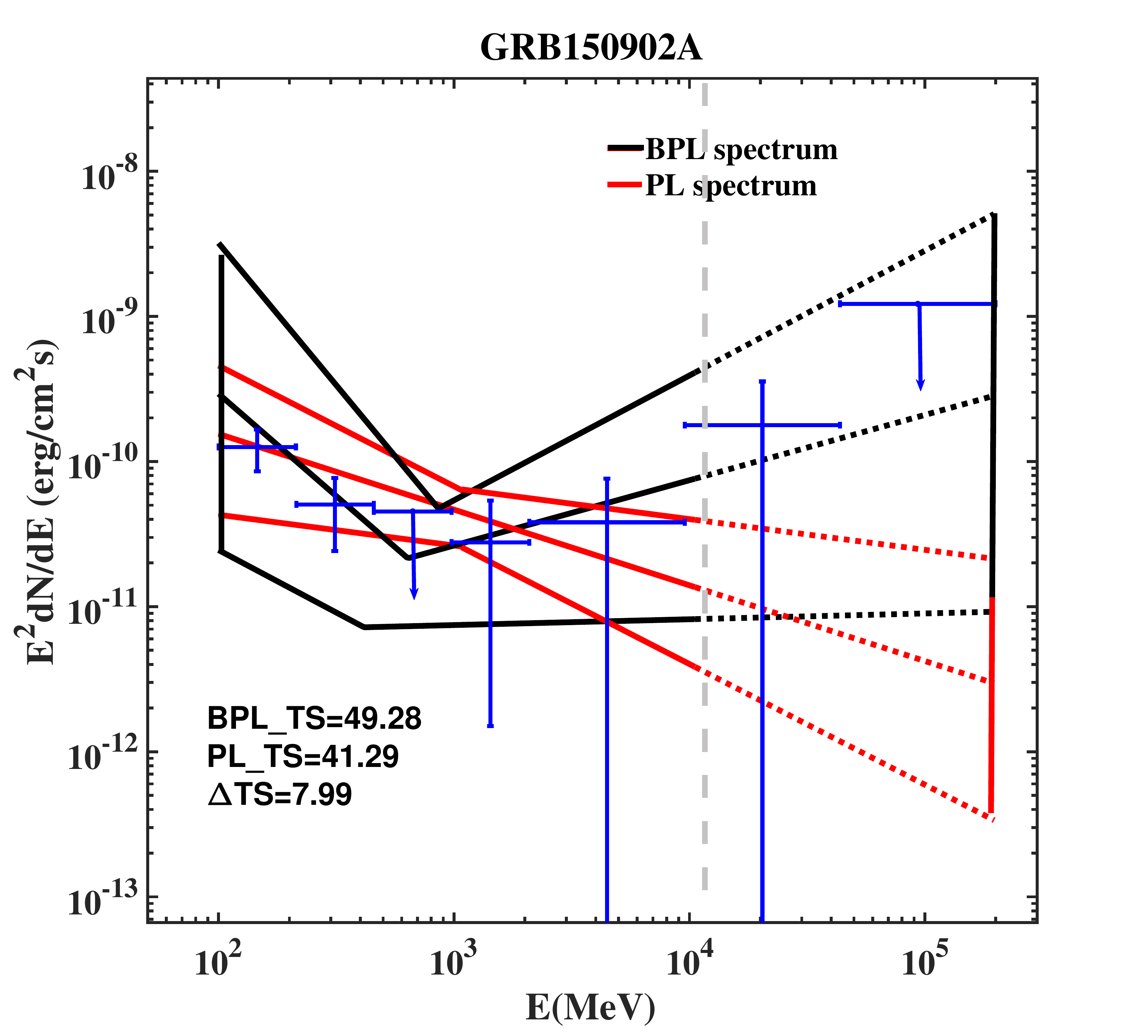}
\includegraphics[width=92mm,height=85mm]{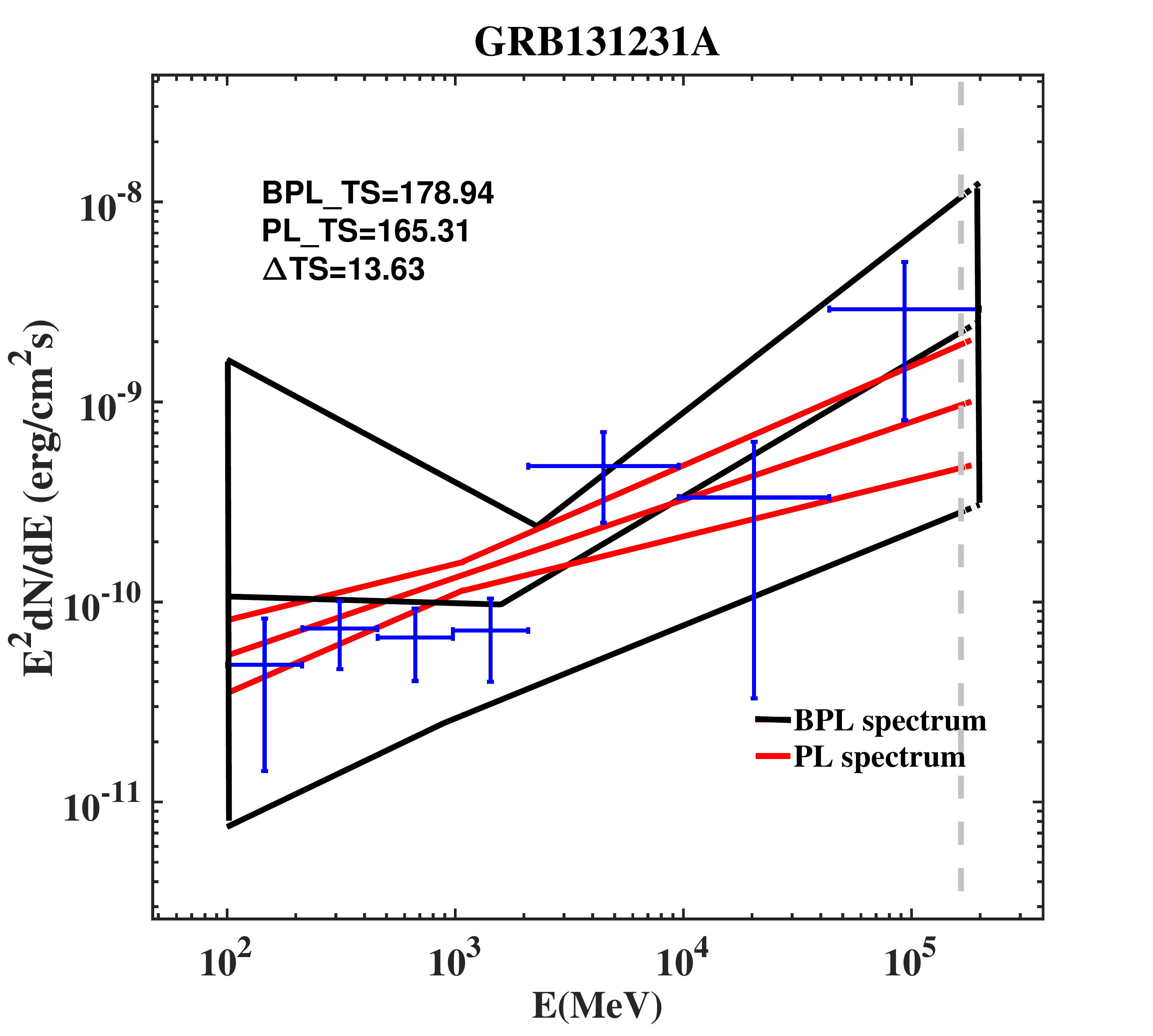}\\
\caption{The 0.1-200~GeV PL and BPL spectra of the 7 GRBs in the $E^2$dN/dE representation, using time range from the 2$\times\,T_\mathrm{90}$ to one day thereafter. Red and black lines show the PL model fits and BPL model fits, respectively. All $\pm 1\sigma$ error contours are propagated from errors on the fit parameters. The vertical, dashed lines indicate the energy of the most energetic photon detected. The spectra are extrapolated above the maximum photon energy to 200 GeV, and are shown as dotted lines. The blue data points are fits with a power-law in individual energy bands. The upper limits are calculated with assuming index=$-3$ (fixed).}
\end{figure*}
\begin{figure*}
\addtocounter{figure}{-1}
\includegraphics[width=92mm,height=85mm]{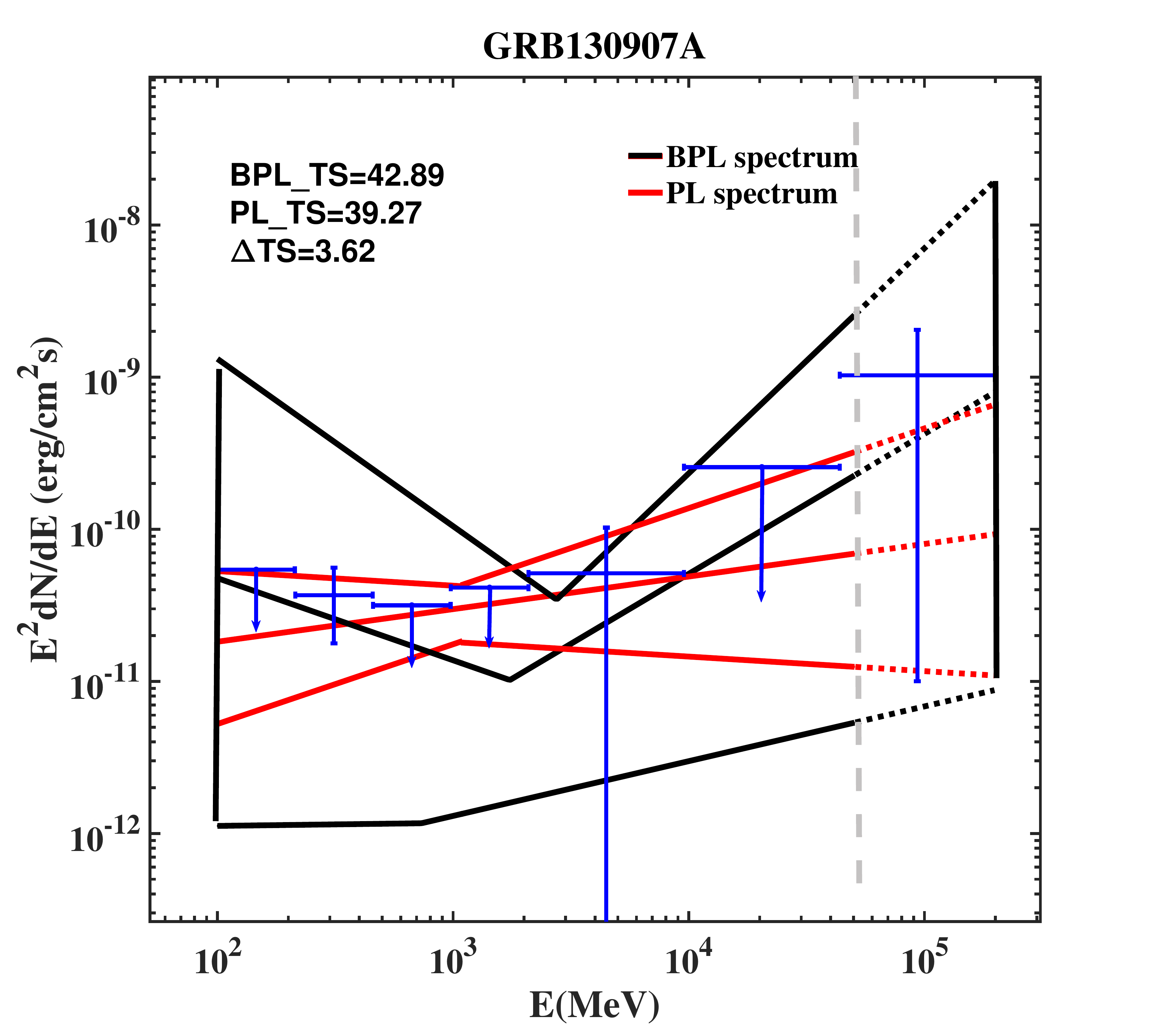}
\includegraphics[width=92mm,height=85mm]{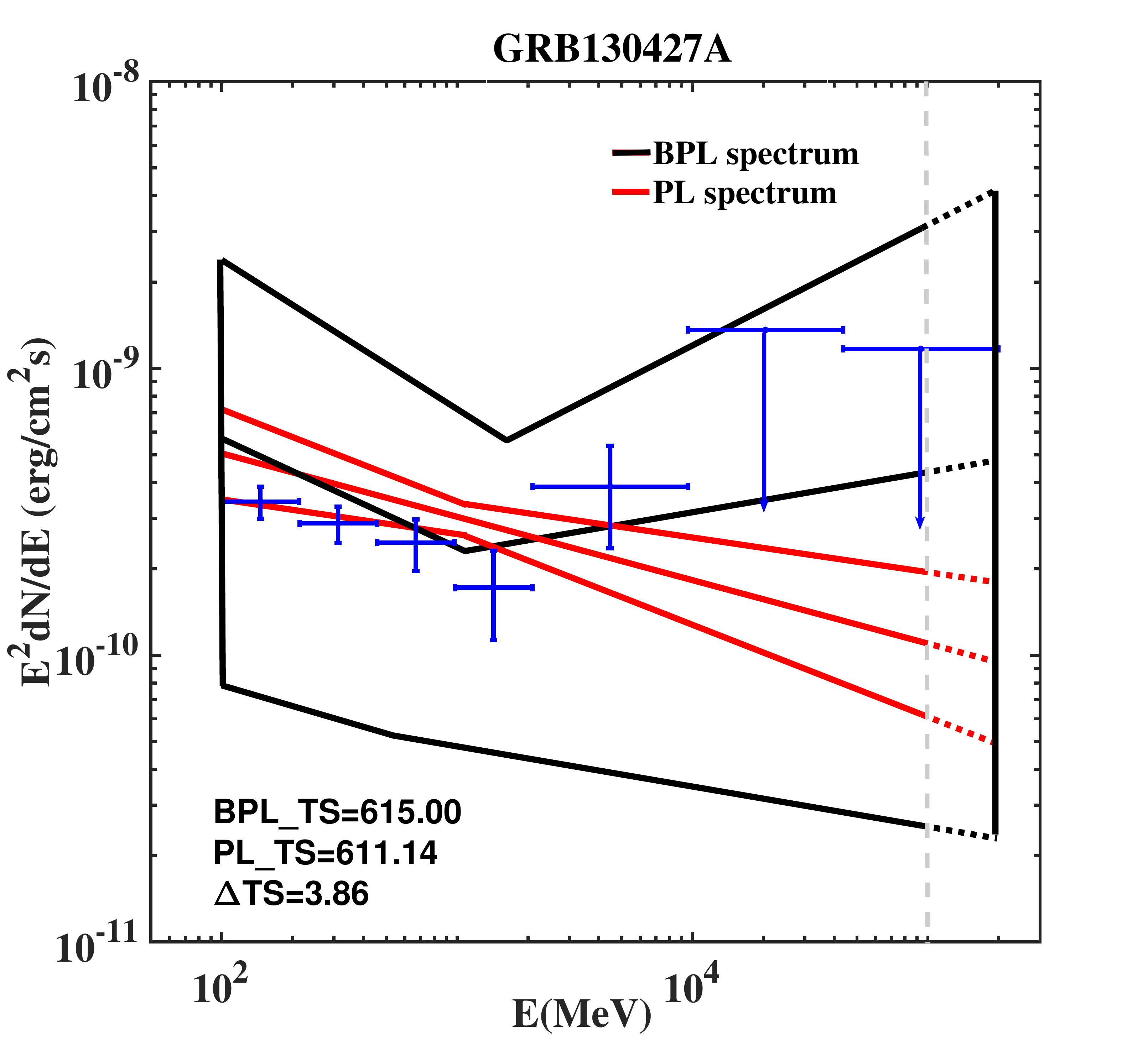}\\\\
\includegraphics[width=92mm,height=85mm]{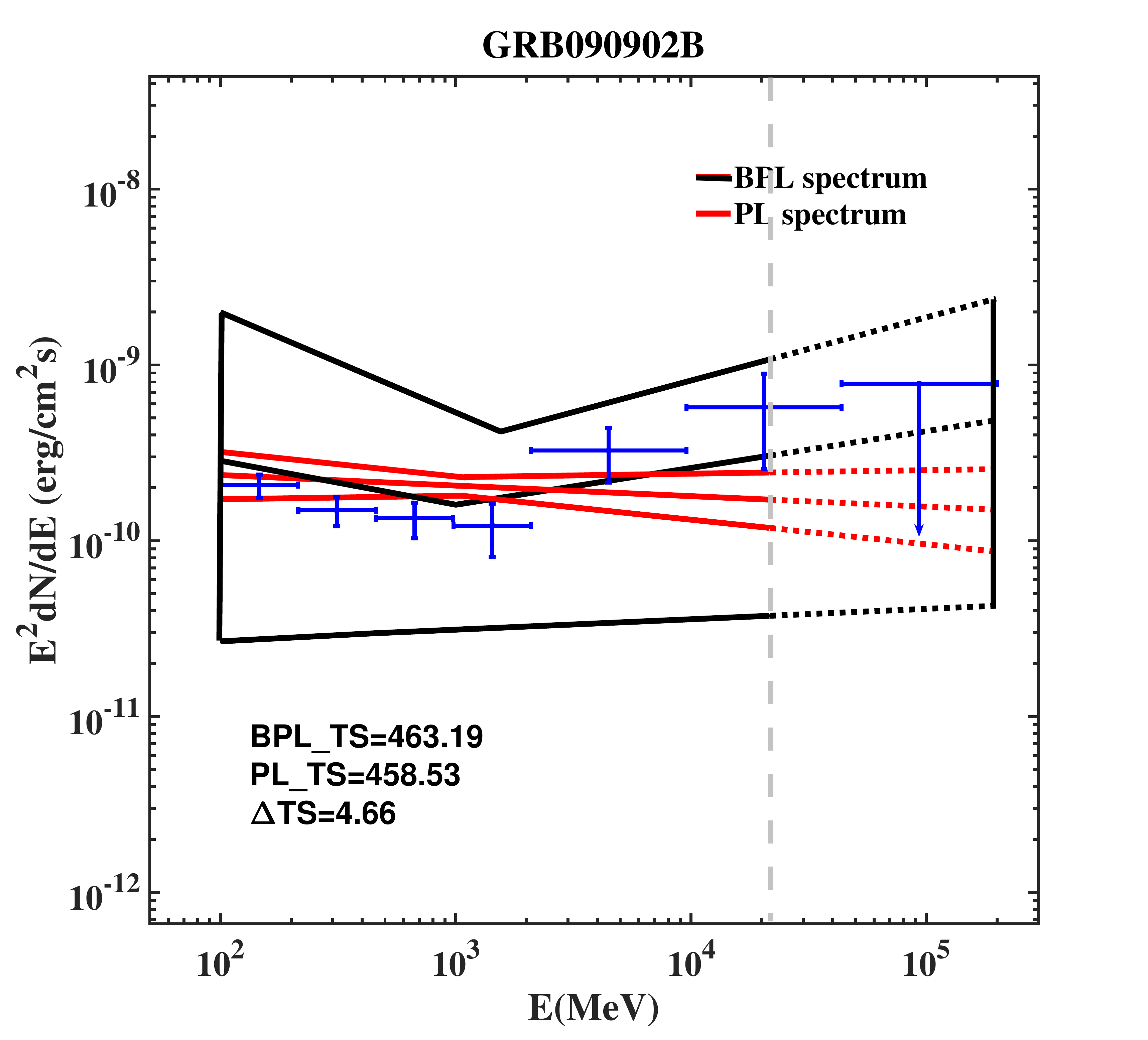}
\caption{continued from previous page}
\label{sample3}
\end{figure*}

\begin{figure*}
\includegraphics[width=60mm,height=48mm]{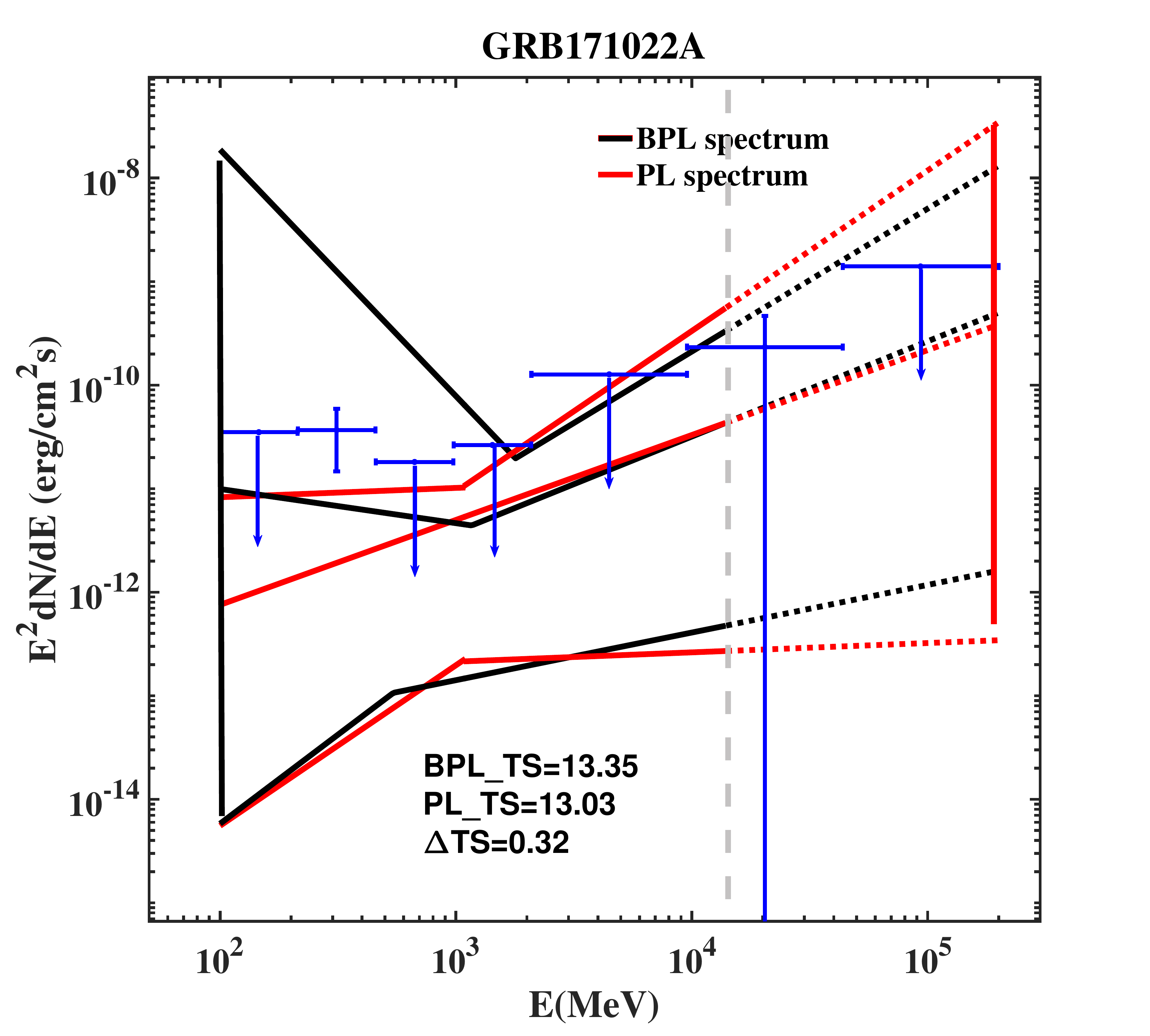}
\includegraphics[width=60mm,height=48mm]{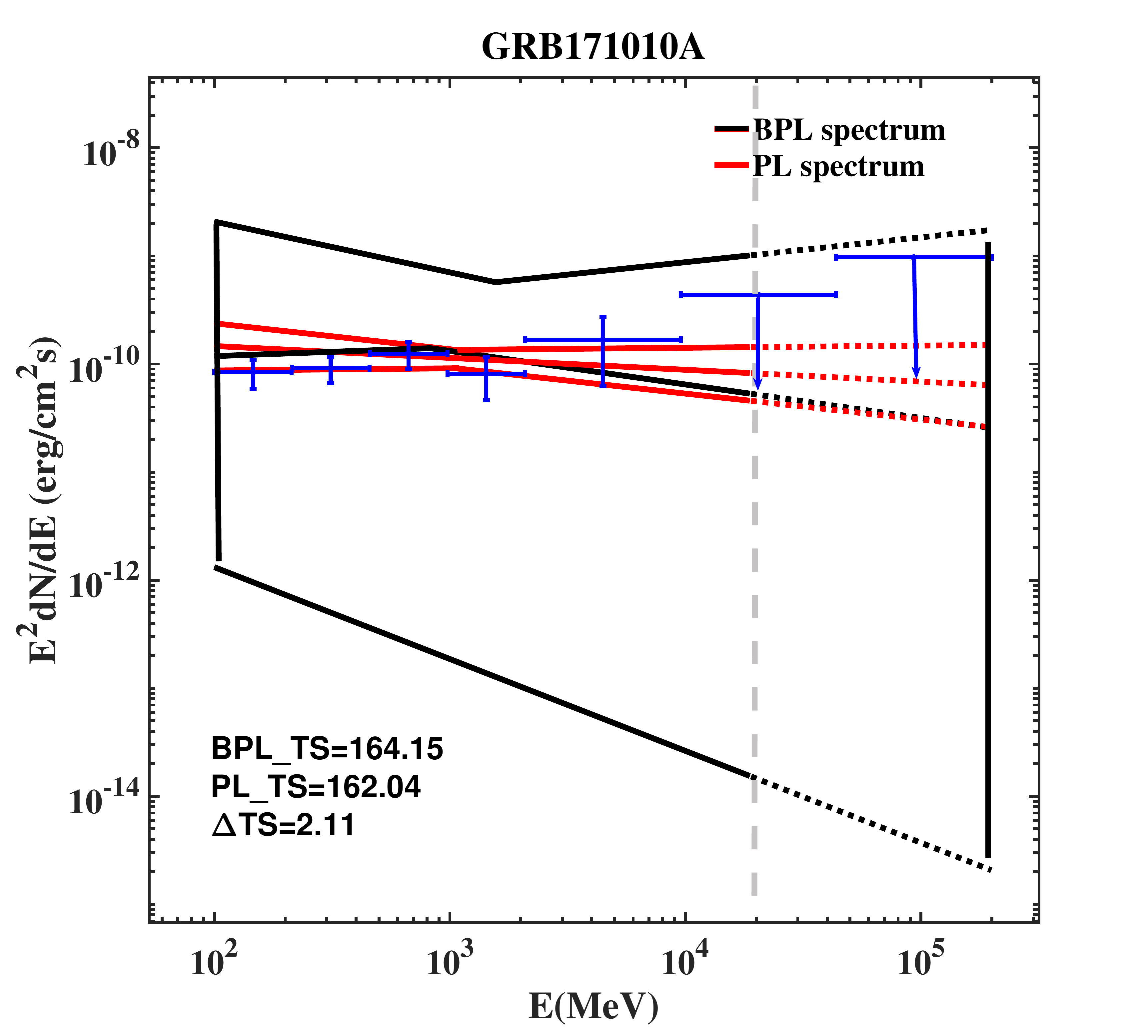}
\includegraphics[width=60mm,height=48mm]{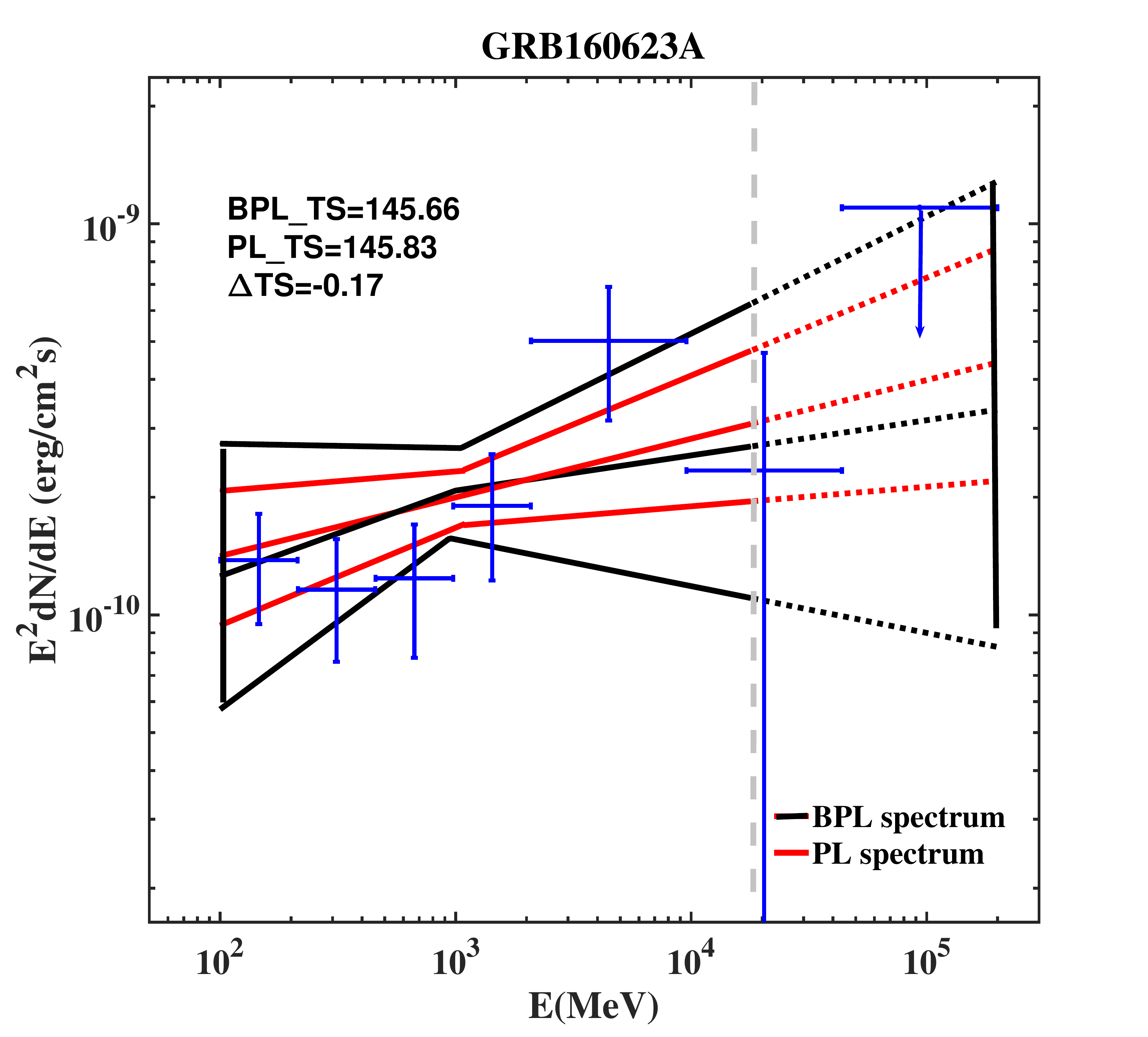}\\\\
\includegraphics[width=60mm,height=48mm]{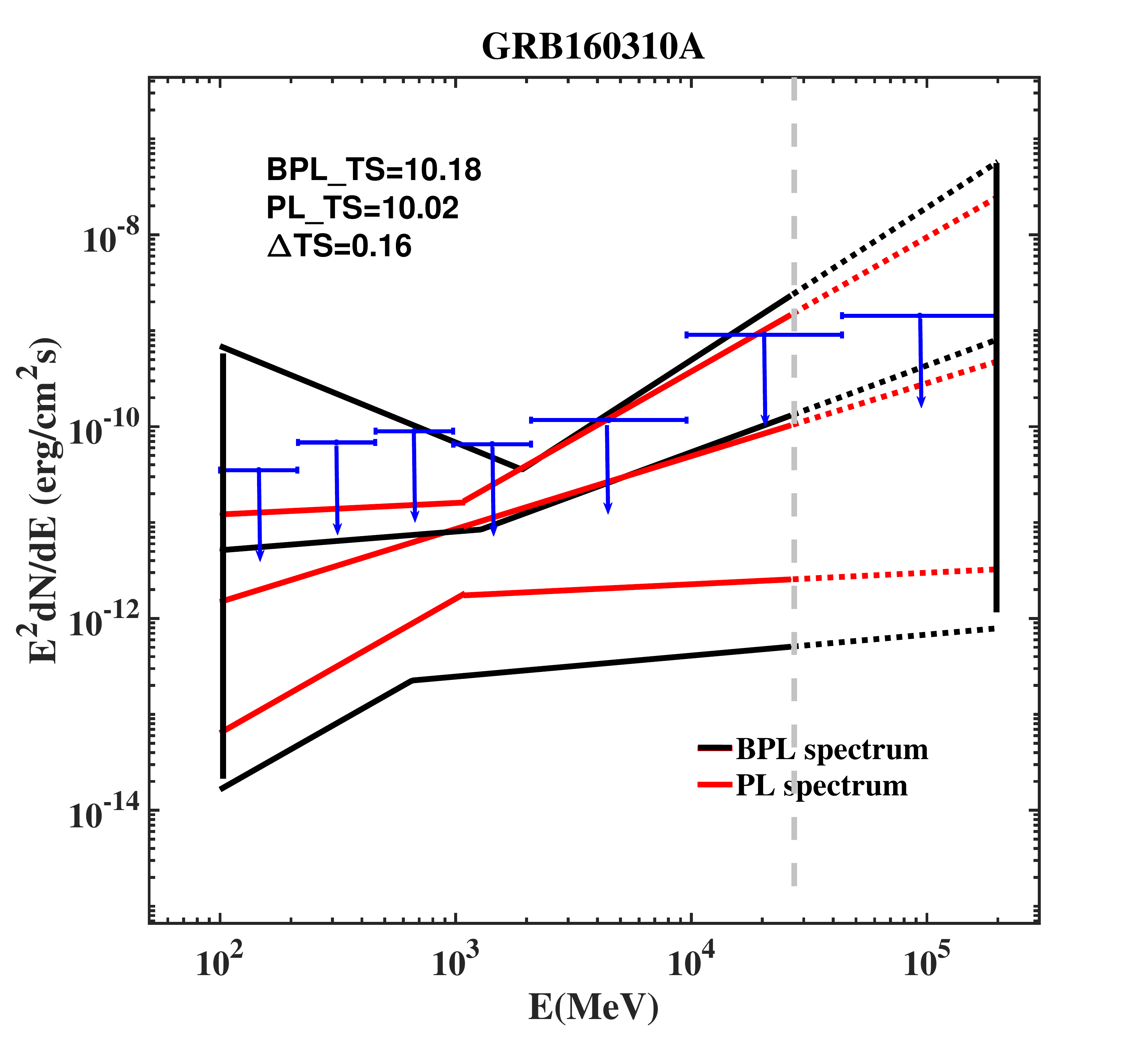}
\includegraphics[width=60mm,height=48mm]{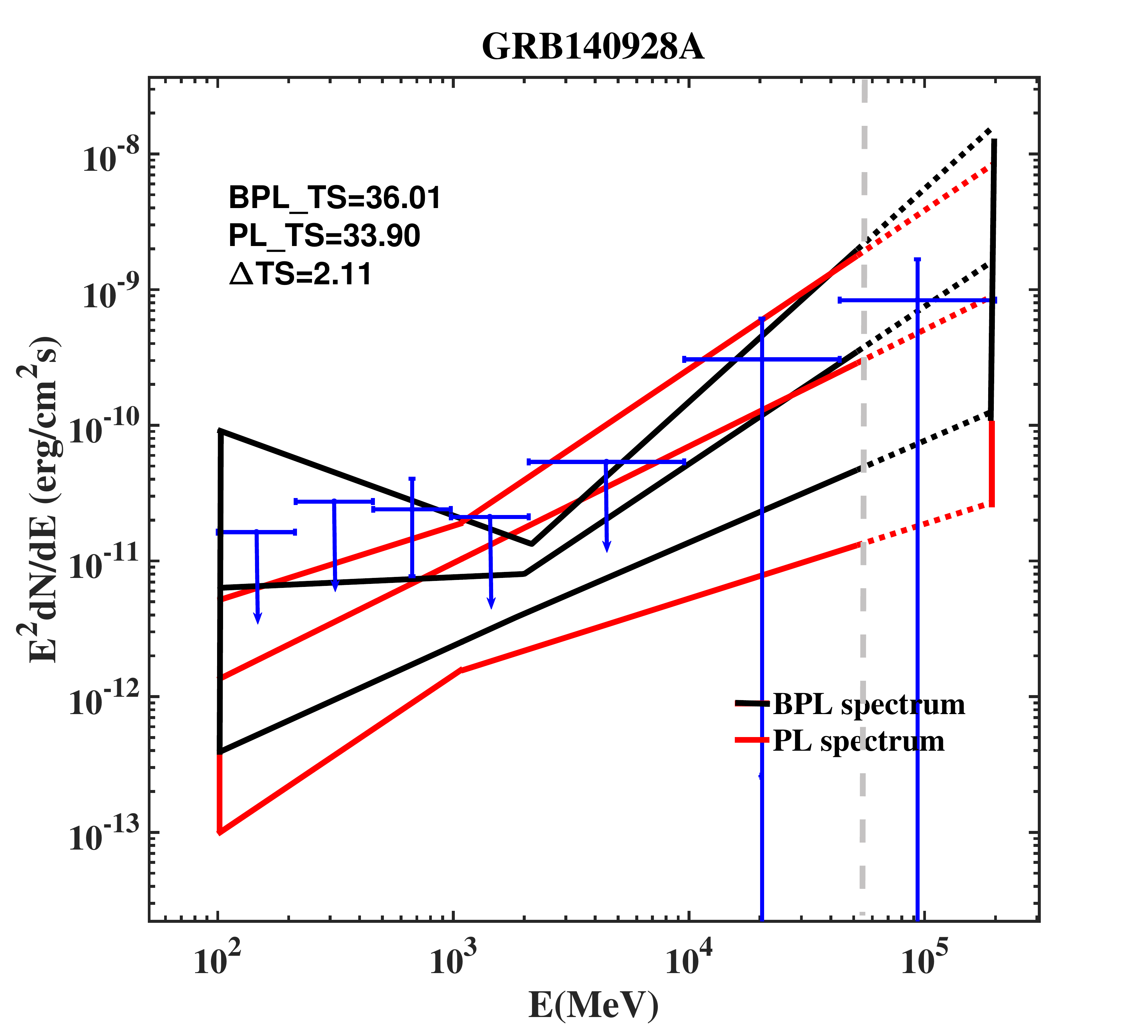}
\includegraphics[width=60mm,height=48mm]{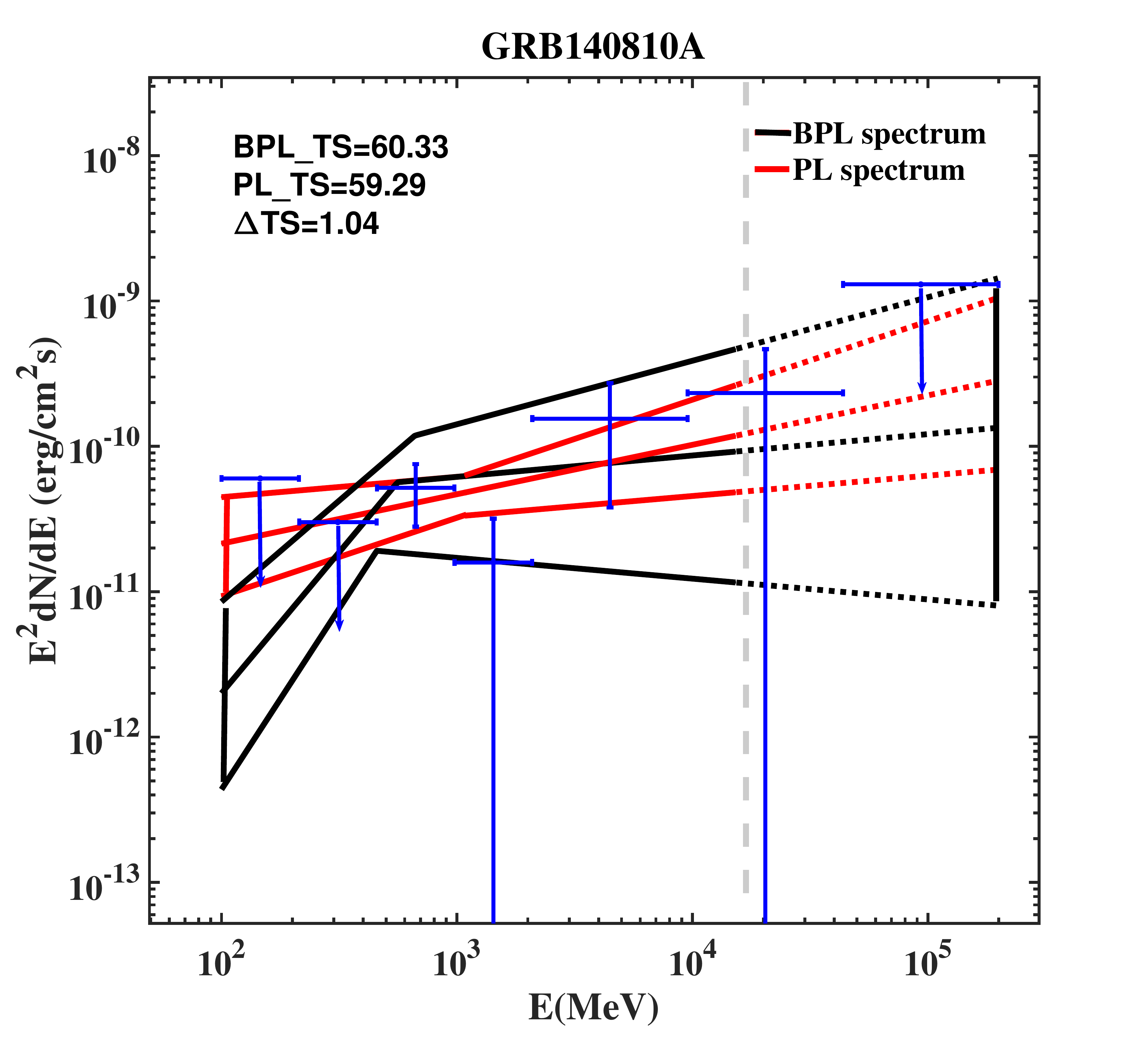}\\\\
\includegraphics[width=60mm,height=48mm]{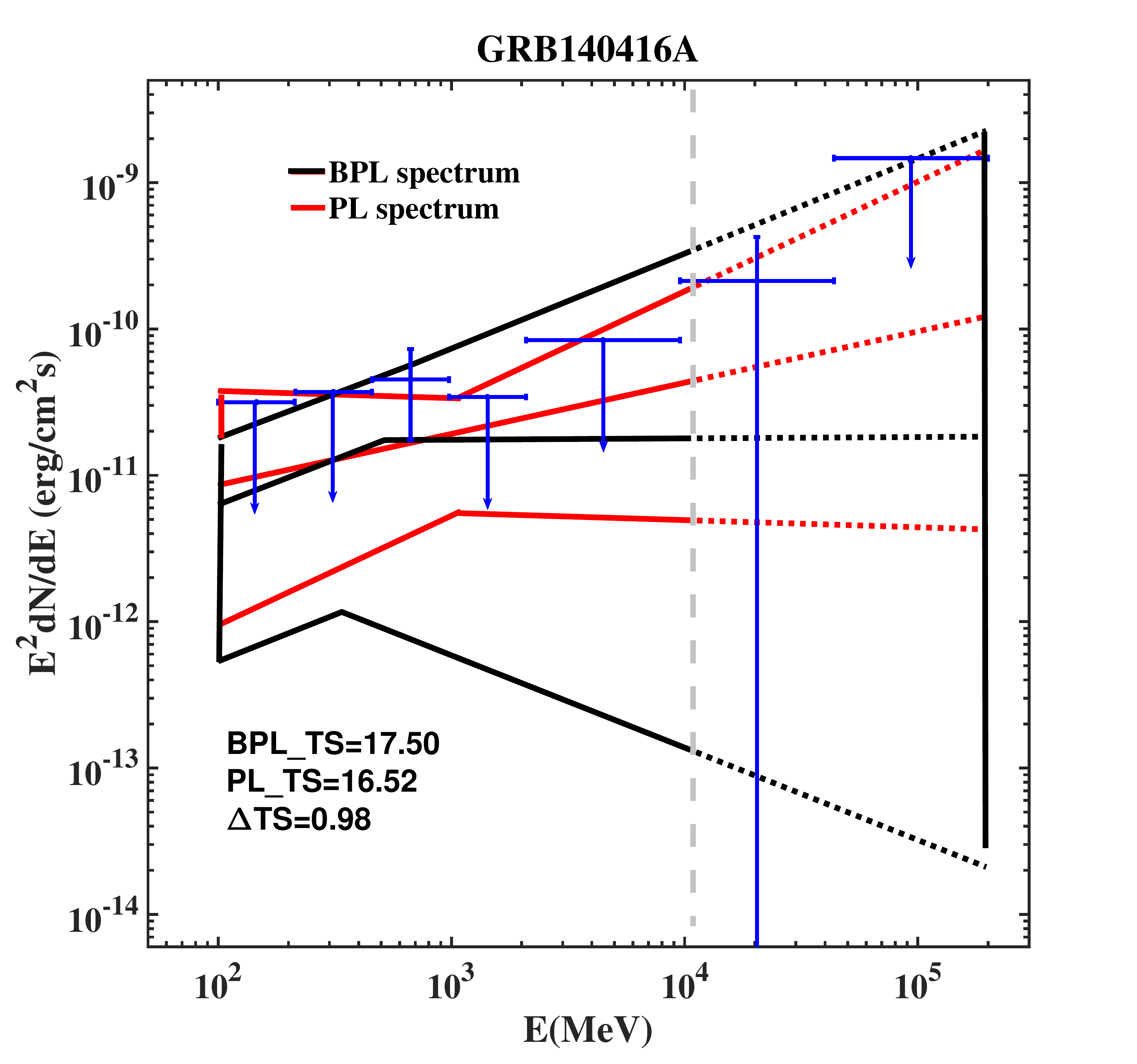}
\includegraphics[width=60mm,height=48mm]{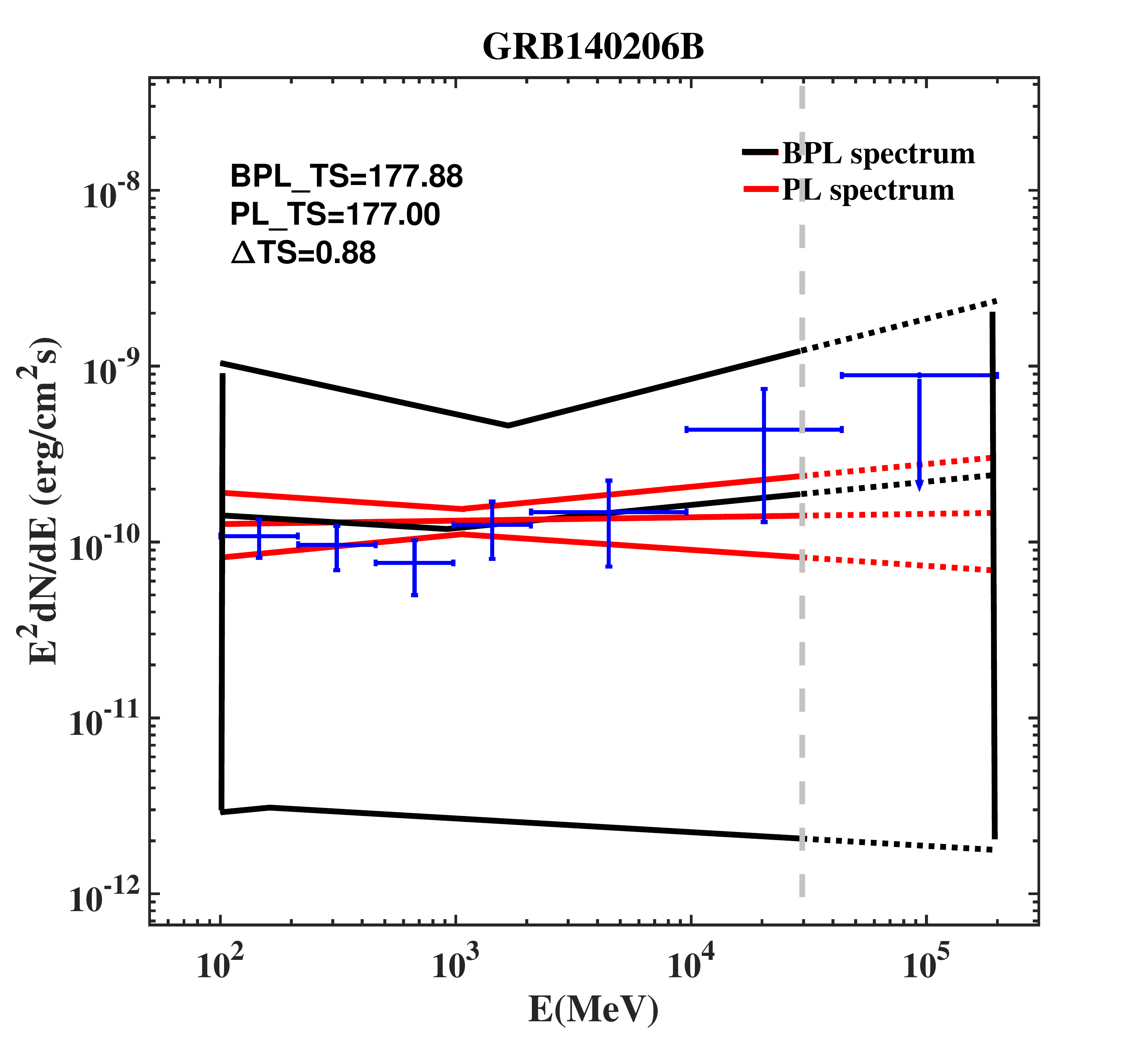}
\includegraphics[width=60mm,height=48mm]{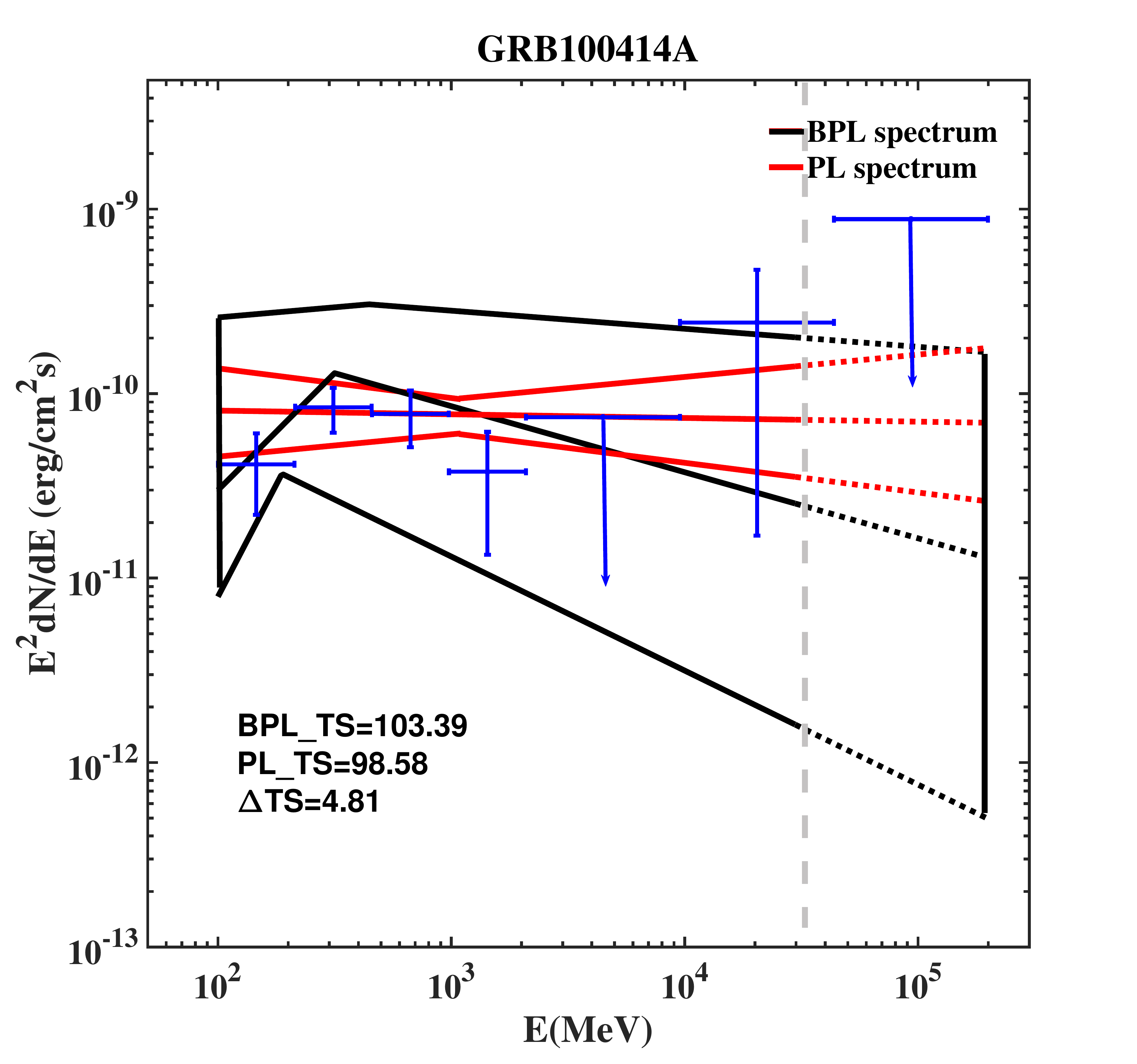}\\\\
\includegraphics[width=60mm,height=48mm]{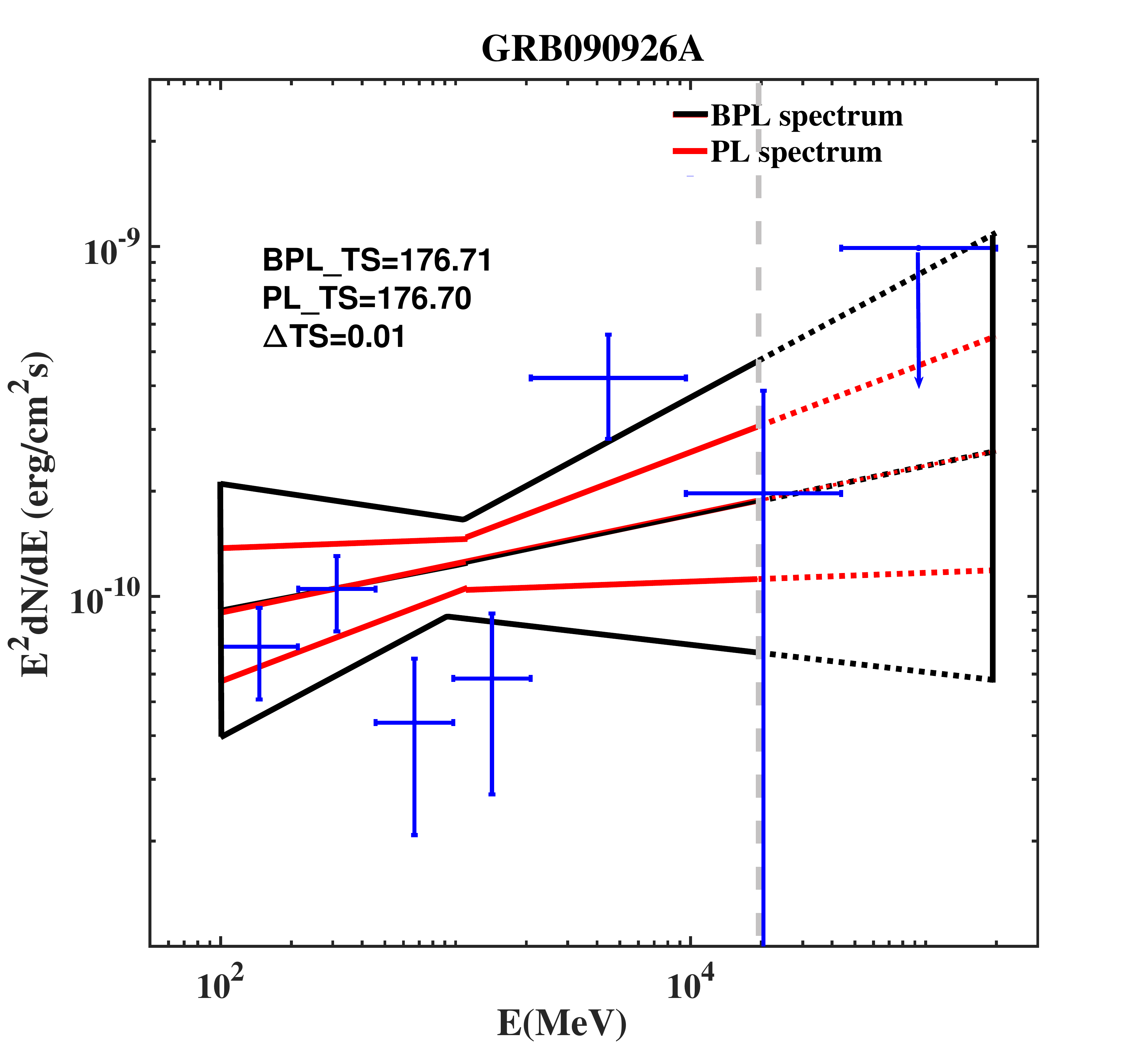}
\includegraphics[width=60mm,height=48mm]{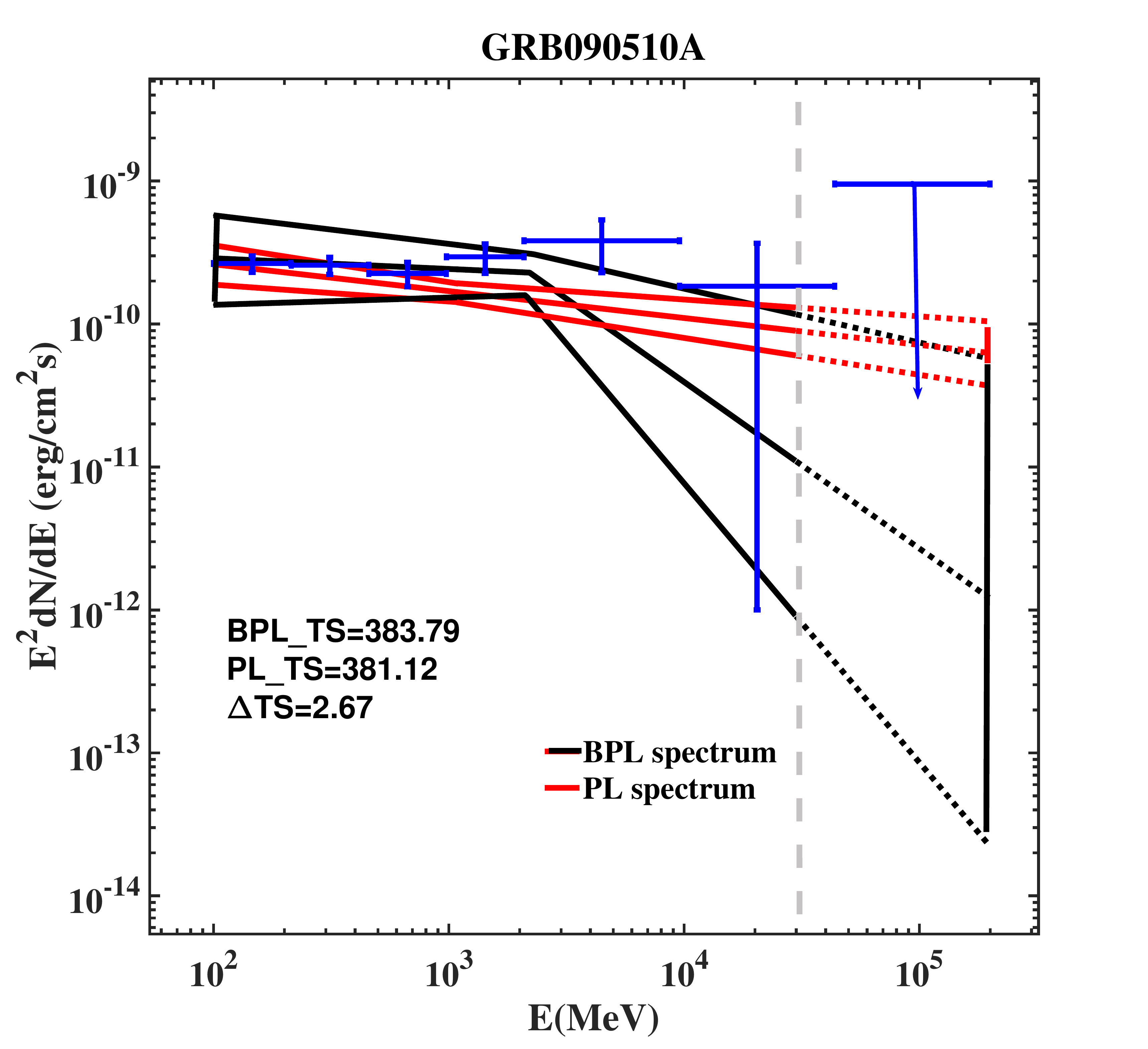}
\caption{The 0.1-200~GeV PL and BPL spectra of the remaining 11 GRBs in the $E^2$dN/dE representation, using time range from the 2$\times\,T_\mathrm{90}$ to one day thereafter. Red and black lines show the PL model fits and BPL model fits, respectively. All the $\pm 1\sigma$ error contours are propagated from errors on the fit parameters. The vertical, dashed lines indicate the energy of the most energetic photon detected, and the spectra are extrapolated above the maximum photon energy to 200 GeV, and are shown as dotted lines. The blue data points are fits with a power-law in individual energy bands. The upper limits are calculated with assuming index=$-3$ (fixed).}
\label{sample2}
\end{figure*}

\begin{figure*}
  \centering
\includegraphics[width=190mm,height=110mm]{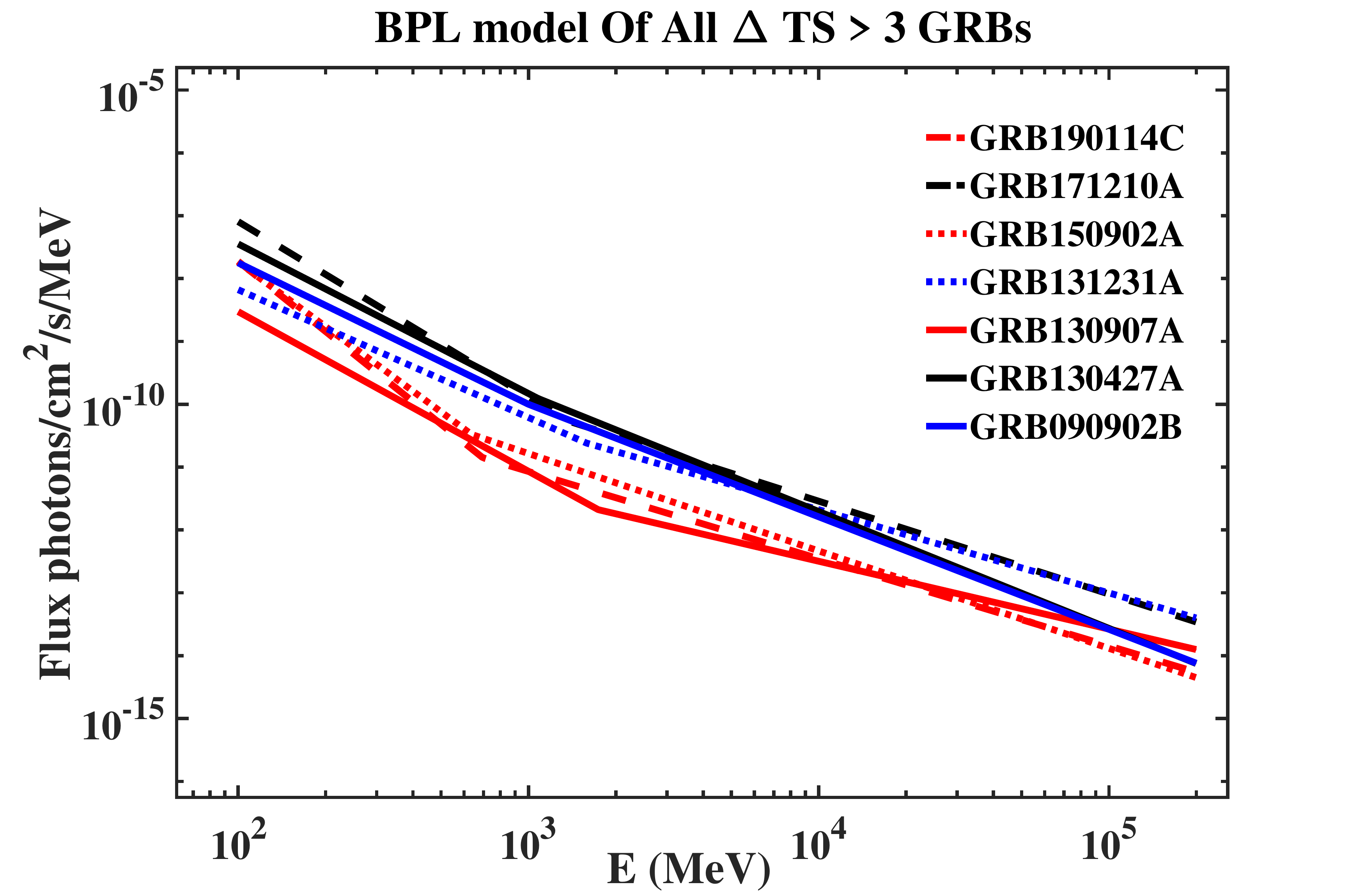}
\caption{The best-fit broken power-law photon spectra of the 7 GRBs, which possibly reveal an upturn in the afterglow spectra and $\Delta$TS$=TS_{BPL}-TS_{PL} >$ 3.}
\label{combined}
\end{figure*}
\begin{figure*}
\includegraphics[width=92mm,height=80mm]{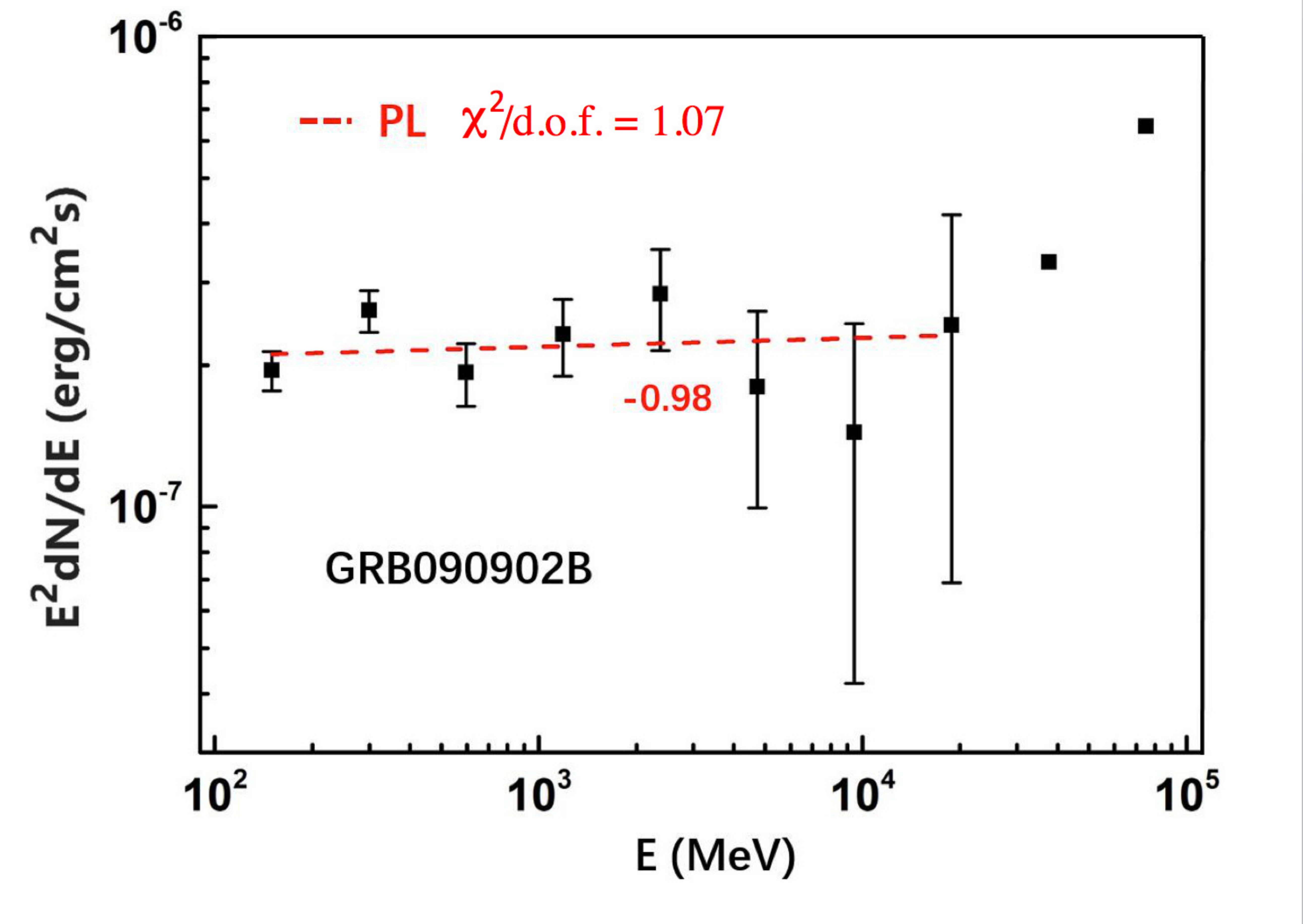}
\includegraphics[width=92mm,height=80mm]{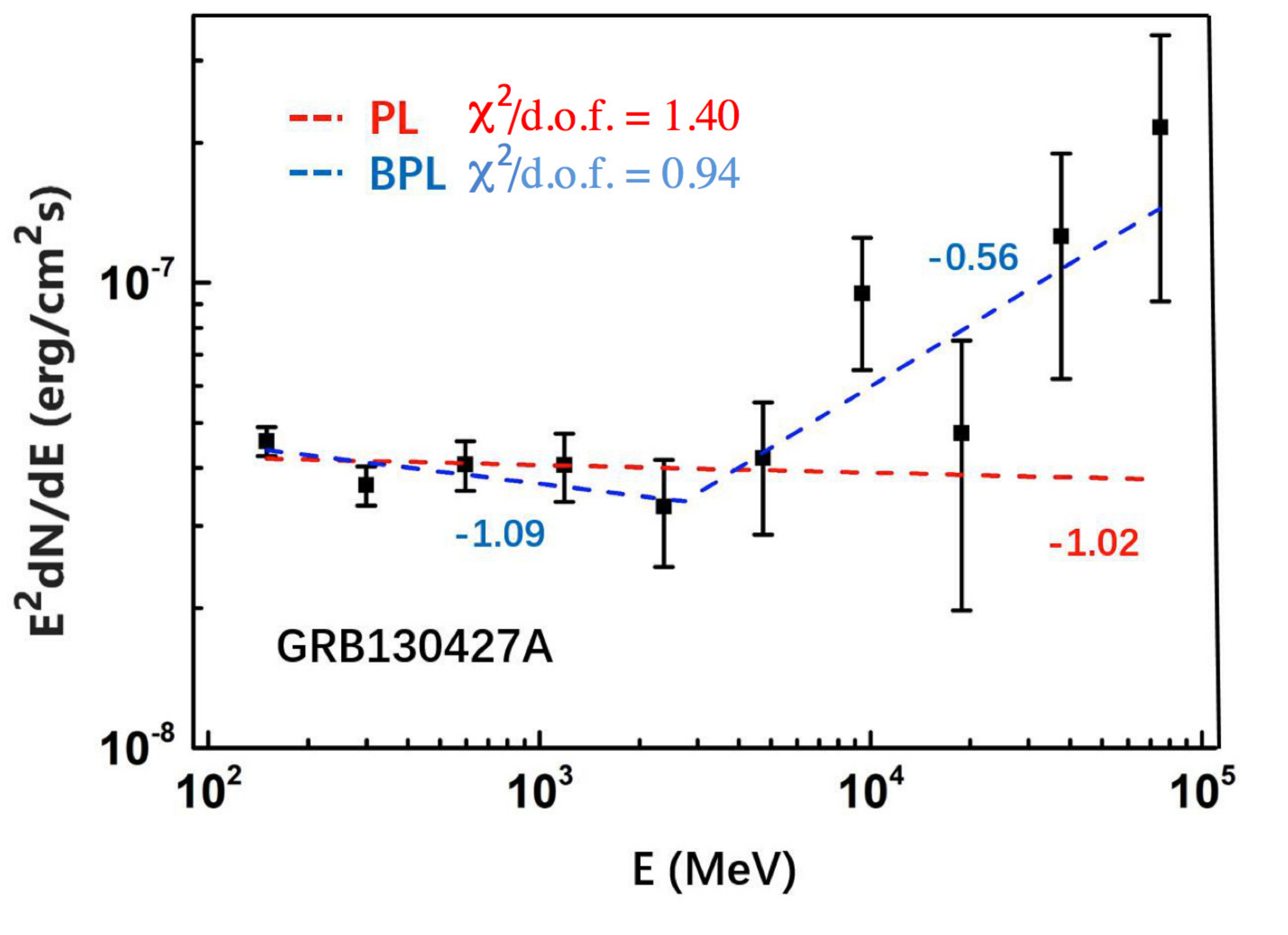}
\includegraphics[width=92mm,height=80mm]{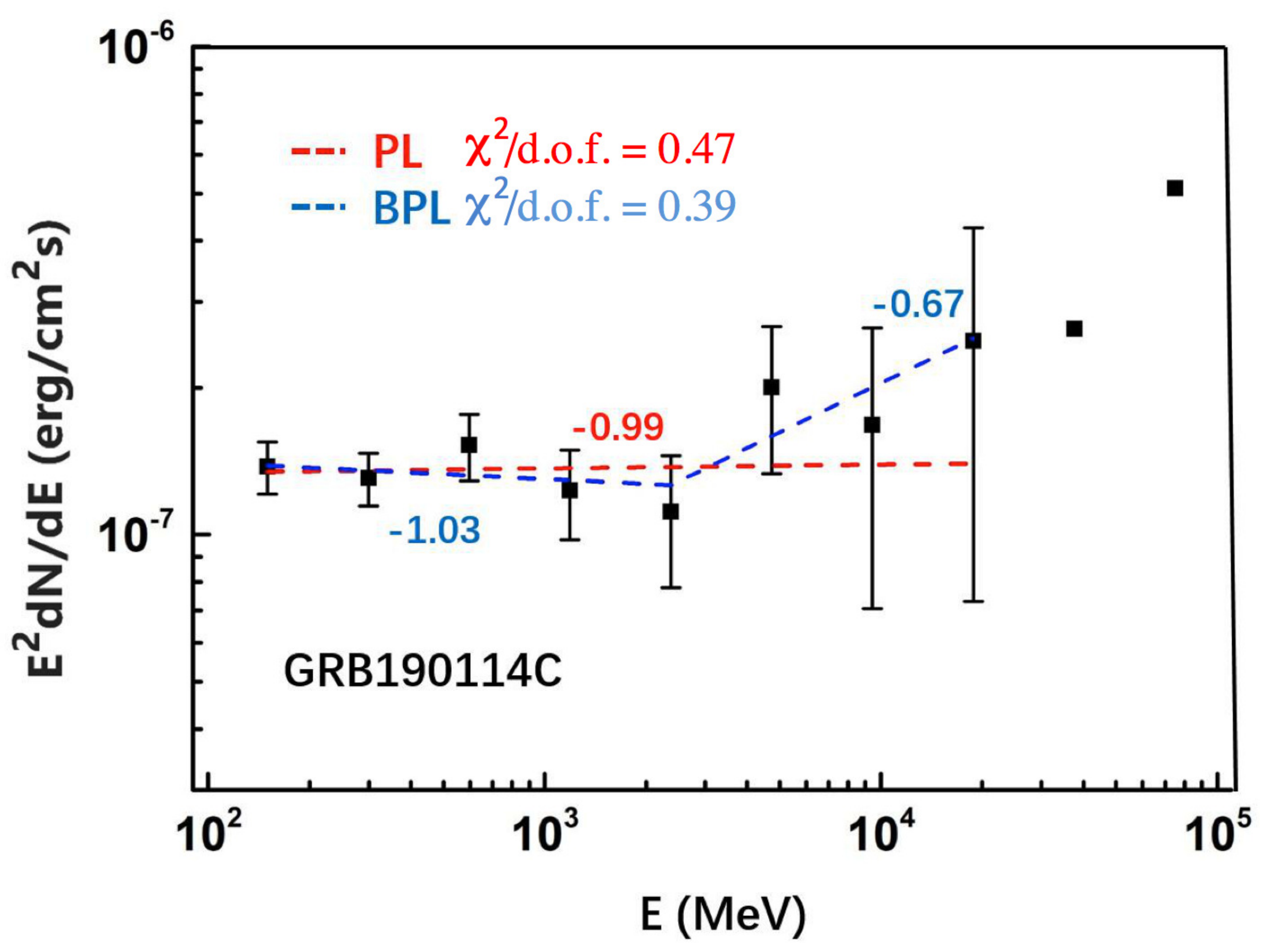}
\caption{The spectra $F$ of GRB~090902B, GRB~130427A and GRB~190114C in the time interval $T_\mathrm{0}$ to 2$\times\,T_\mathrm{90}$. The single power-law fits are made starting from 100 MeV and up to where measurements can be fit. Data (squares) without error bars are upper limits, which are calculated assuming $\Gamma$=$-3$ (fixed). The dash lines are the fits of the data points with a PL (red) or BPL (blue) model. The numbers next to the model fits are the index of the spectra $F$ in each energy interval. } %All 3 GRBs show an upturn in higher energy range. They are good evidences for inverse-Compton above the spectral break.}
\label{sample4}
\end{figure*}

\begin{figure*}
\centering
  \includegraphics[width=140mm,height=80mm]{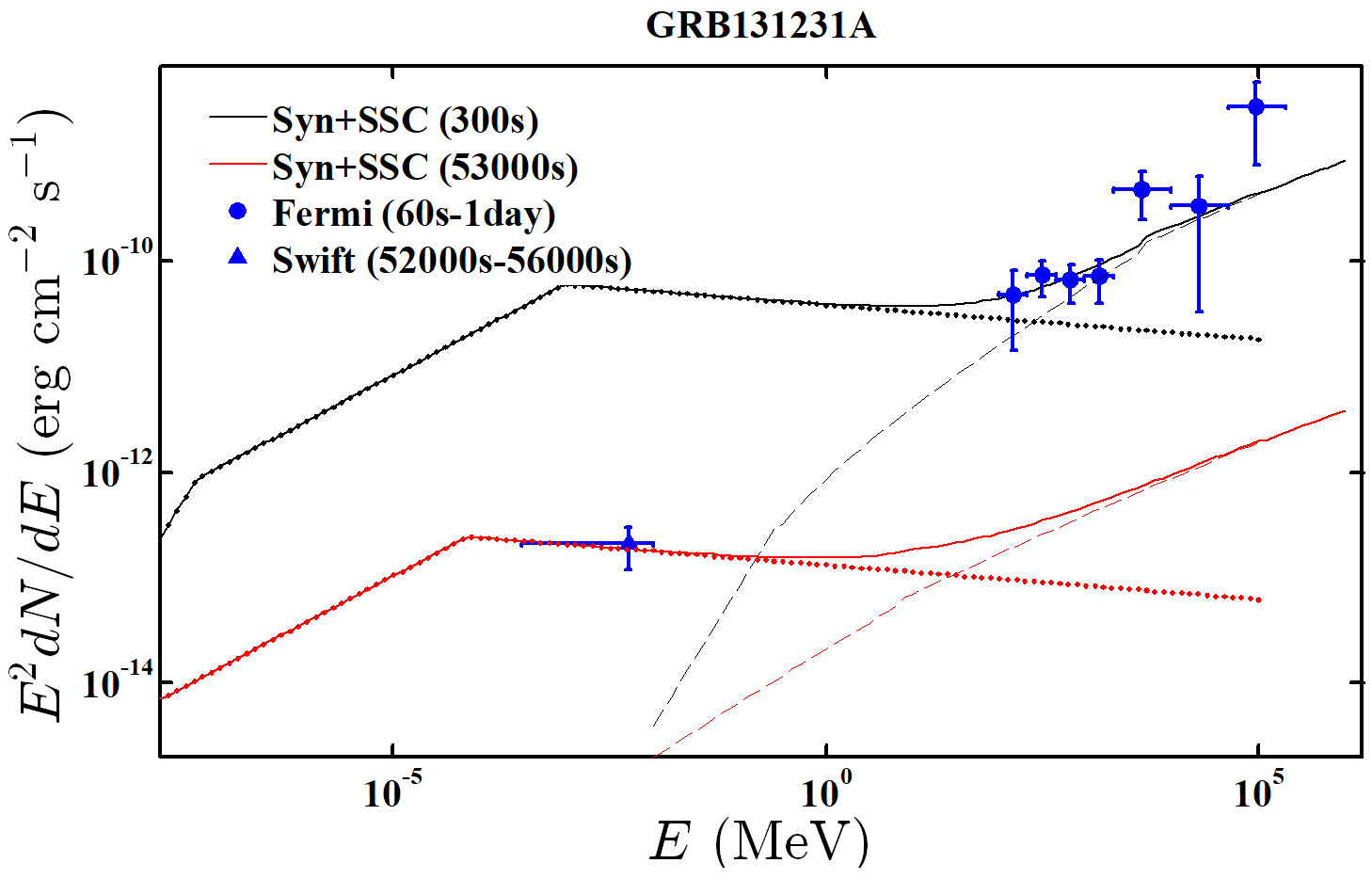}
  \caption{\small{Modelling the broadband spectra of GRB~131231A in the time intervals 60s-1day and 52000-56000s. Thick black curves represent the theoretical spectra of synchrotron plus SSC corresponding to slow cooling in the external forward shock scenario. Dotted line and dashed line represents the synchrotron and SSC component, respectively. The adopted parameters are $\epsilon_e$=0.16, $\epsilon_B$=0.00001, $p$=2.13, and $n$=1 $\rm cm^{-3}$}.}
  \label{modeling}
\end{figure*}

\begin{figure*}
  \centering
\includegraphics[width=140mm,height=108mm]{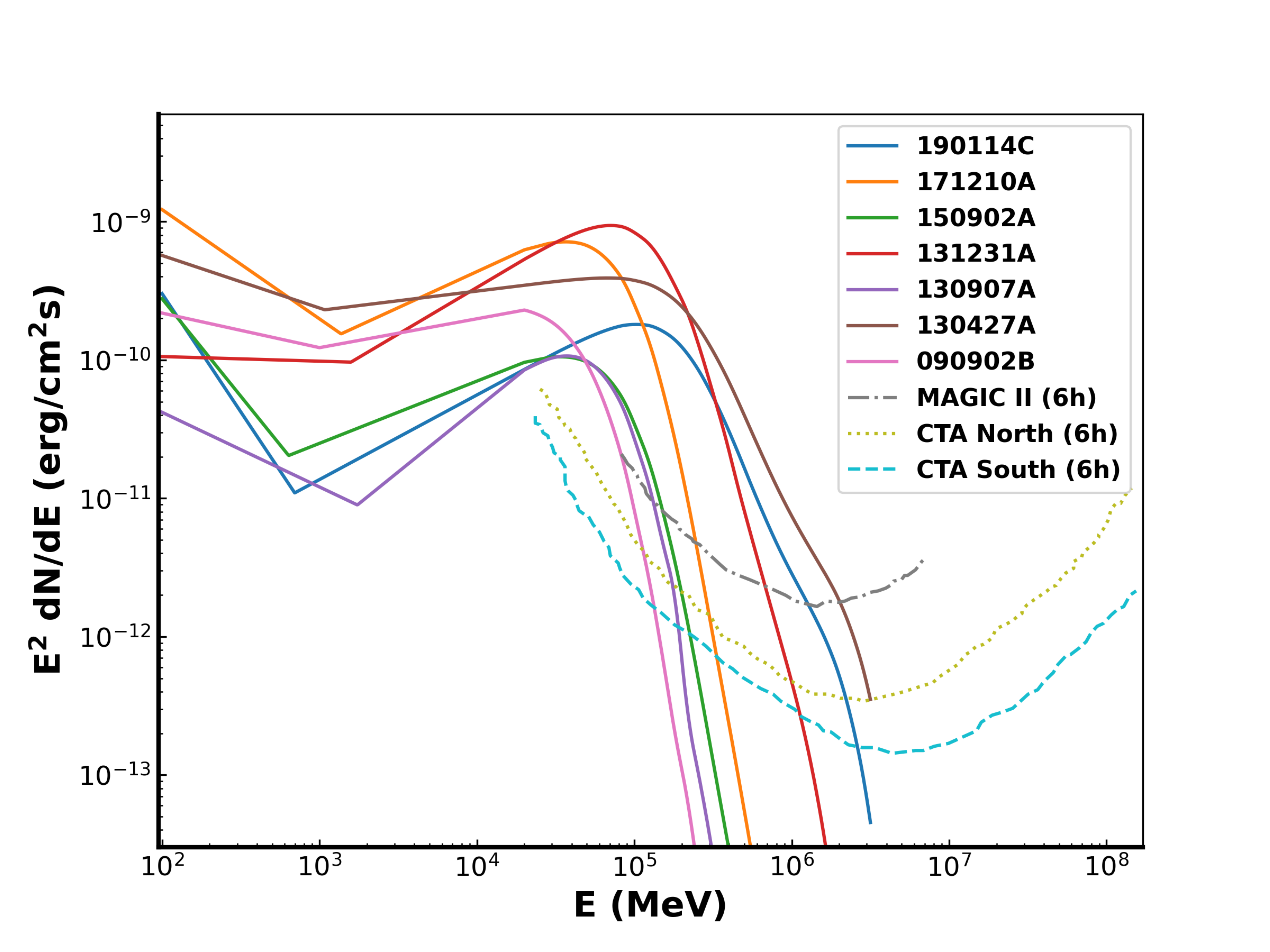}
\caption{The broken power-law energy spectra of the 7 GRBs and the 6-hour sensitivity of CTA and MAGIC. The time interval of the 7 GRBs are 2$\times\,T_\mathrm{90}$ to 24h. We assume the high-energy index $\beta$ of the intrinsic spectra extends up to 10~TeV. }
\label{sensitivity}
\end{figure*}

\end{document}